\documentclass{article}

 \usepackage[main, final]{neurips_2026}

\usepackage[utf8]{inputenc} 
\usepackage[T1]{fontenc}    
\usepackage{url}            
\usepackage{booktabs}       
\usepackage{amsfonts}       
\usepackage{nicefrac}       
\usepackage{microtype}      
\usepackage{xcolor}         

\usepackage{xspace}
\usepackage{enumitem}
\usepackage{graphicx}
\usepackage{caption}
\usepackage{wrapfig}
\usepackage{amsmath}
\usepackage{multirow}
\usepackage{colortbl}
\usepackage{makecell,array}
\usepackage{fontawesome5}

\usepackage{subcaption}
\definecolor{cvprblue}{rgb}{0.21,0.49,0.74}

\definecolor{tableblue}{RGB}{229,242,255}
\definecolor{tablepurple}{RGB}{219,223,249}
\definecolor{tablepeach}{RGB}{238,224,242}

\definecolor{tablegray}{RGB}{242,242,242}

\definecolor{textblue}{RGB}{33,70,138}
\definecolor{textpurple}{RGB}{70,76,150}
\definecolor{textgray}{RGB}{127,127,127}

\newcommand{\stdblue}[1]{$_{\textcolor{textblue}{\tiny \pm #1}}$}
\newcommand{\stdpurple}[1]{$_{\textcolor{textpurple}{\tiny \pm #1}}$}
\newcommand{\stdgray}[1]{$_{\textcolor{textgray}{\tiny \pm #1}}$}

\usepackage[colorlinks=true,
            citecolor=cvprblue,
            linkcolor=cvprblue,
            urlcolor=cvprblue]{hyperref}
\setlist[itemize]{topsep=-1pt, itemsep=1pt, parsep=0pt, partopsep=0pt,left=1pt}

\newcommand{\linkitem}[3]{%
  {\hypersetup{hidelinks}%
   \href{#1}{\mbox{%
     \raisebox{-0.15ex}{%
       \resizebox{!}{9.5pt}{\textcolor{black}{#2}}%
     }%
     \hspace{0.35em}%
     {\fontsize{9.5pt}{12pt}\selectfont
      \textbf{\textcolor{black}{#3}}}%
   }}}%
}

\title{Focus, Align, and Diffuse: Time-Series-Aware Keyframe Diffusion for Cardiac Dynamic Synthesis from Sparse Observations}

\author{%
  Junkai Liu$^1$,\;\; 
  Haofan Wu$^1$,\;\; 
  Nay Aung$^2$,\;\; 
  Joao A. C. Lima$^3$, \\
  \textbf{Steffen E. Petersen}$^2$,\;\; 
  \textbf{Le Zhang}$^{1\dagger}$ \\[3pt]
    $^1$ School of Engineering, University of Birmingham \\
    $^2$ William Harvey Research Institute, Queen Mary University London \\
    $^3$ Division of Cardiology, Johns Hopkins University School of Medicine \\[3pt]
    \texttt{l.zhang.16@bham.ac.uk} \\[8pt]
    \linkitem{https://cardiofad.github.io/}{\faGlobe}{Project Page}\qquad
    \linkitem{https://github.com/JK-Liu7/CardioFAD}{\faGithub}{GitHub}
}

\begin{document}

\maketitle

{\let\thefootnote\relax\footnotetext{$\dagger$: Corresponding author}}

\begin{abstract}
Dense cardiac dynamics are essential for assessing cardiac function, yet in practice clinicians often observe only extremely sparse measurements. However, recovering full dynamics from such sparsity is a highly ill-posed problem due to the lack of temporal cues. While auxiliary time-series signals like ECG provide physiological guidance, effectively exploiting them is challenging due to the cross-modal representation gap and phase misalignment. In this work, we present \textbf{CardioFAD}, a \textbf{Focus--Align--Diffuse} paradigm that reformulates sparse cardiac dynamic synthesis as a \emph{recoverability--alignability anchor discovery} problem. Our core insight is that resolving extreme temporal ambiguity requires a compact set of anchor states that are both sufficient to recover the visual trajectory and reliable for ECG--visual alignment. We therefore learn motion-salient keyframes under two complementary principles: \emph{trajectory sufficiency}, which preserves trajectory-critical intra-modal motion content, and \emph{anchor alignment}, which promotes consistent inter-modal correspondence at these anchors. We implement the \textbf{Focus} and \textbf{Align} modules optimized by the two aspects. \textbf{Diffuse} then performs two-stage generation by first synthesizing keyframes and subsequently interpolating the remaining frames. Extensive experiments on cardiac MRI and echocardiography videos demonstrate superior visual quality, temporal coherence, and improved functional consistency under the extreme one-shot regime. 


\end{abstract}    
\section{Introduction}

\begin{figure*}[t]
\centering
\includegraphics[width=1.00\linewidth]{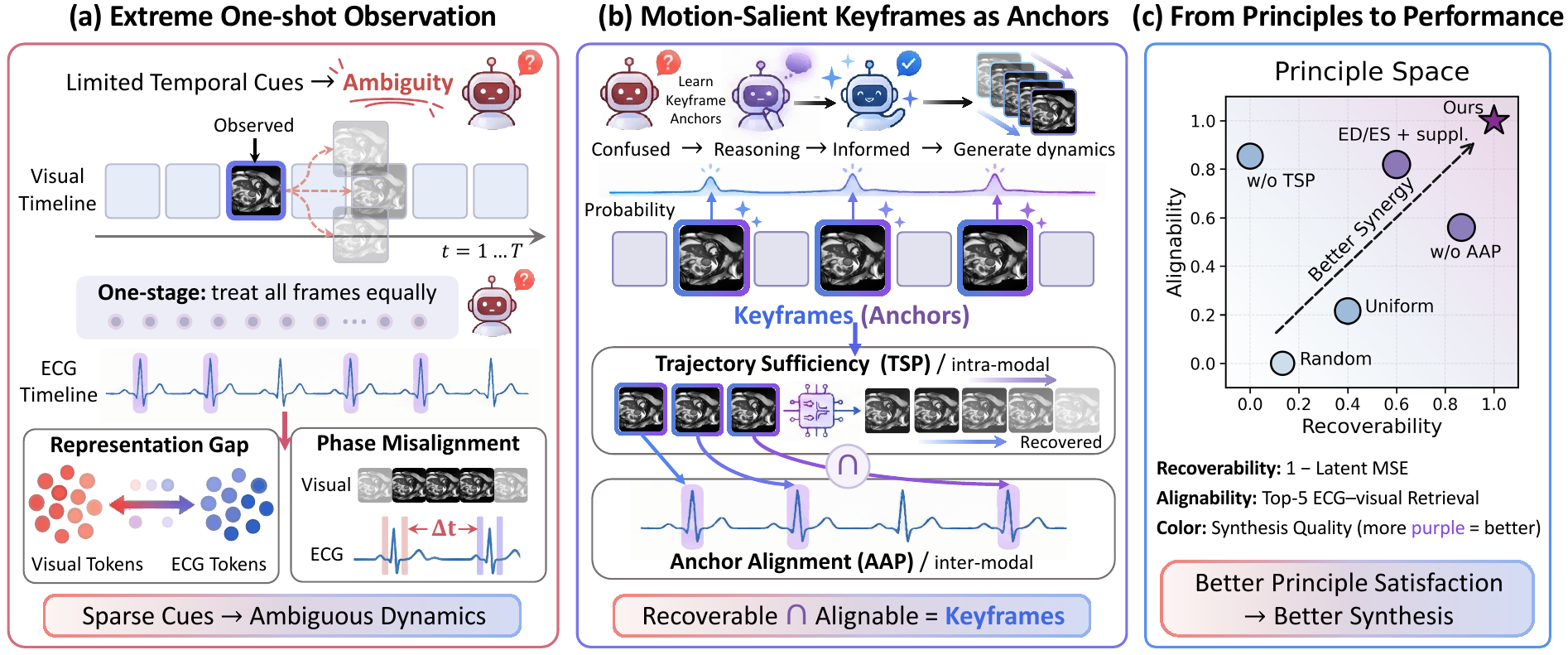}
\vspace{-12pt}
\caption{\textbf{(a) Challenge:} Extreme sparsity causes temporal ambiguity, while ECG suffers from cross-modal misalignments. \textbf{(b) Solution:} We reformulate the task as a recoverability--alignability anchor discovery problem. \textbf{(c) Principle-to-Performance:} Our learned keyframes satisfy both principles, yielding superior synthesis quality.}
\vspace{-18pt}
\label{teaser}
\end{figure*}

Dynamic cardiac imaging modalities, such as cardiac MRI (CMR) and echocardiography (ECHO), are widely used to capture cardiac motion and monitor disease progression for assessing heart function~\cite{doi:10.1016/j.jacc.2024.05.050, doi:10.1148/rg.325115098, doi:10.1161/CIRCRESAHA.116.309128}. Many clinically relevant cues are characterized by motion over the cardiac cycle~\cite{qiao2025personalized, 10797672}, yet acquiring sequences (3D+t CMR or 2D+t ECHO videos) is expensive and time-consuming, and clinicians often have access to only a sparse observation (e.g., a single frame or a single slice)~\cite{liu2025sagcnet, XIA2021101812}. Despite progress in generating entire sequences ~\cite{li2025echopulse, Liu_2024_ACCV, chen2025end}, the problem remains underexplored, especially under extreme one-shot sparsity due to its \textit{ill-posed} nature. 

The main challenge of this task lies in the fact that vastly different spatiotemporal trajectories can explain the same limited evidence~\cite{zhang2019missing, deb2025unsupervised, LE2025100720, 10988690, Kim_2024_CVPR, ma2025cardiacflow, Zhang_2026_CVPR}. Given only an extremely sparse visual observation (e.g., a single frame or slice), the model lacks sufficient motion cues to anchor the global cardiac-state trajectory (Fig.~\ref{teaser}a). This makes rapid contraction and relaxation phases highly underdetermined and prone to temporal drift or motion inconsistency~\cite{You_2025_ICCV, PUNG2026103308, 10.1145/3691344}. Intuitively, not all time points are equally informative for characterizing the trajectory, thus selecting salient keyframes is critical to preserving essential cardiac motion content~\cite{wang2025keyvid, zhu2025focus, Bae_2025_ICCV}. Randomly or uniformly sampled keyframes are often suboptimal, while using predefined phases such as end-diastole (ED) and end-systole (ES) as anchors is too coarse to capture richer dynamics beyond ED/ES across diverse pathologies and acquisition protocols. Meanwhile, although ECG is a natural source of auxiliary cues, translating it into reliable visual guidance is non-trivial. ECG provides phase-related electrophysiological cues but does not specify anatomy or appearance, and its correspondence to imaging states can vary across subjects, introducing a cross-modal representation gap and phase misalignment~\cite{pan2026tracing, selivanov2025global, chen2025versatile, li2025echopulse} (Fig.~\ref{teaser}a). In this context, existing one-stage conditional diffusion models treat each frame equally, and may ignore the weak conditioning, over-smooth salient motion, or hallucinate temporally inconsistent dynamics~\cite{guo2024sparsectrl, 10.1145/3680528.3687652, Burgert_2025_CVPR, Jeong_2025_CVPR}. Hence, an effective solution should learn motion-salient anchors and align ECG with visual dynamics in a temporally aware manner, so that the ECG condition can reliably constrain generation.

Motivated by the above analysis, we ask the central question: \textit{How can we synthesize dense cardiac dynamics from an extremely sparse observation while effectively leveraging auxiliary time-series signals?} Our \textbf{key insight} is to reformulate this ill-posed task as a \emph{recoverability--alignability anchor discovery} problem. Rather than treating all frames equally or relying on stronger global conditioning, the model should identify a compact set of motion-salient states that are both (i) sufficient to recover the intra-modal cardiac trajectory and (ii) reliable for ECG--visual alignment (Fig.~\ref{teaser}b). Such recoverable-and-alignable anchors reduce temporal ambiguity and turn auxiliary time-series signals into more effective guidance for dense dynamic synthesis.

\begin{figure*}[t]
\centering
\includegraphics[width=1.00\linewidth]{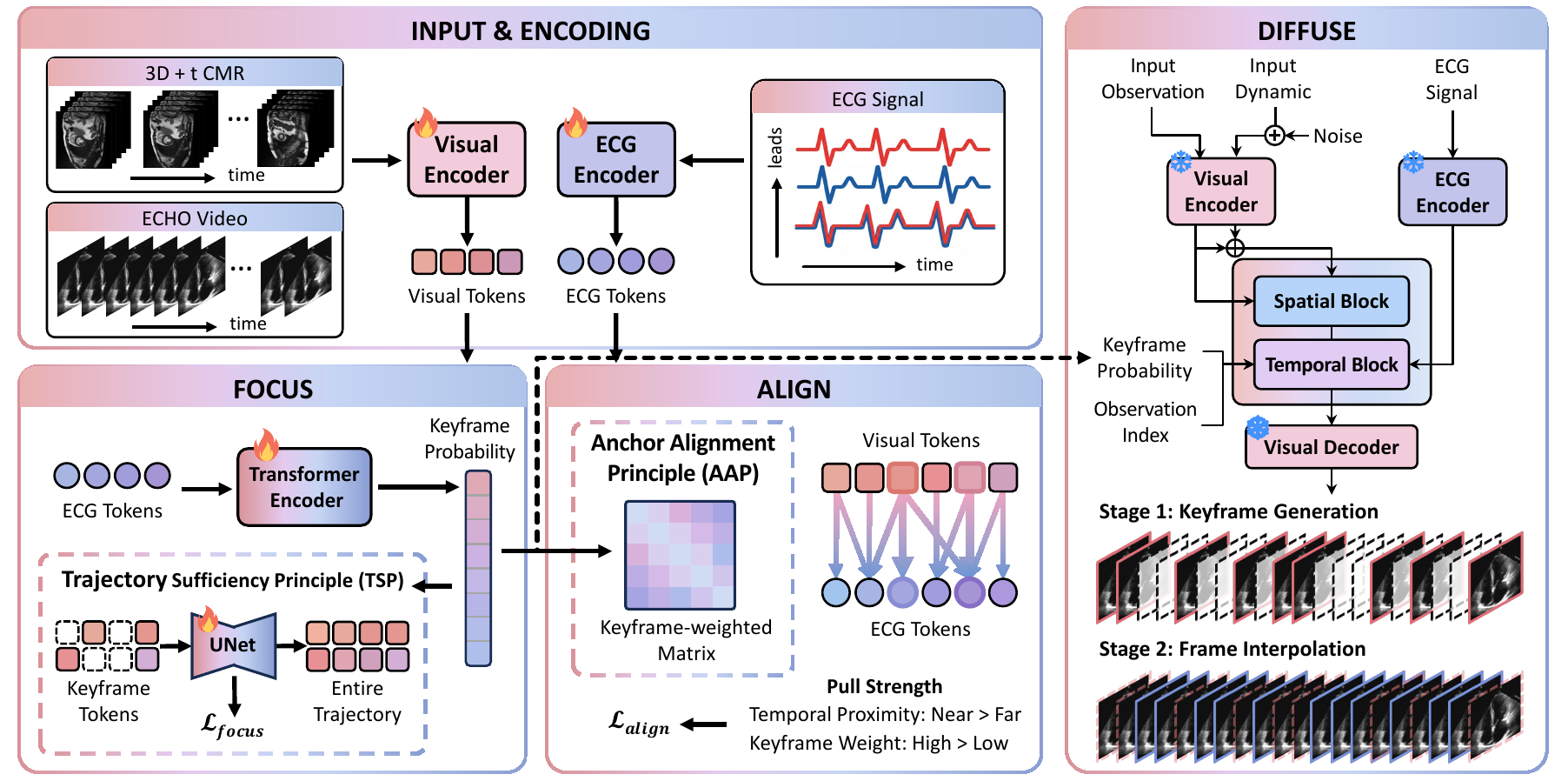}
\vspace{-16pt}
\caption{\textbf{Overview of CardioFAD.} Our pipeline contains three stages: \textbf{(i) Focus:} A tokenizer predicts keyframe probabilities, optimized by the TSP to ensure the selected anchors can reconstruct full intra-modal dynamics. \textbf{(ii) Align:} Guided by the AAP, a keyframe-weighted matrix synchronizes visual and ECG latents, establishing keyframes as cross-modal anchors. \textbf{(iii) Diffuse:} A video DiT performs two-stage synthesis, first generating the  keyframes, then interpolating the remaining frames.}
\vspace{-16pt}
\label{framework}
\end{figure*}

To exploit this recoverability--alignability anchor discovery view, we introduce \textbf{CardioFAD}, a \textbf{\underline{F}ocus--\underline{A}lign--\underline{D}iffuse} paradigm for time-series-conditioned cardiac dynamic synthesis from sparse observations, as shown in Fig.~\ref{framework}. Instead of treating keyframes as predefined landmarks or merely adding stronger global conditioning, CardioFAD learns anchors that satisfy two principle-level criteria: (i) \textbf{trajectory sufficiency}, where informative keyframes should preserve enough intra-modal evidence to reconstruct the full cardiac trajectory; and (ii) \textbf{anchor alignment}, where these keyframes should provide reliable ECG--visual correspondence at motion-salient phases. Concretely, \textbf{Focus} learns sparse yet informative keyframes from input ECG in a self-discovered manner by enforcing the \emph{Trajectory Sufficiency Principle \textbf{(TSP)}}, encouraging the model to allocate capacity to a distinct anchor set that best explains the remaining dynamics. \textbf{Align} implements the \emph{Anchor Alignment Principle \textbf{(AAP)}} with temporally aware, keyframe-weighted ECG-visual alignment, thus turning ECG into reliable guidance. Finally, \textbf{Diffuse} performs two-stage diffusion by first generating anchor keyframes and then interpolating the remaining frames, yielding temporally coherent dynamics and improved functional fidelity under extreme sparsity (Fig.~\ref{teaser}c). Our contributions can be summarized as follows:
 \begin{itemize}
    \item We reformulate one-shot cardiac dynamic synthesis as a \emph{recoverability--alignability anchor discovery} problem, and propose CardioFAD, a keyframe-to-sequence diffusion framework that learns anchors sufficient for visual trajectory recovery and reliable for ECG--visual alignment.
    \item We instantiate this reformulation with a novel trajectory-aware tokenizer guided by TSP and AAP, enabling self-emergent keyframe discovery and keyframe-weighted cross-modal alignment under a unified principle-driven design.
    \item Comprehensive experiments on both 3D+t CMR and 2D+t ECHO benchmarks demonstrate our superior performance in generative quality and cardiac functional consistency under the extreme one-shot regime. Furthermore, principle-targeted diagnostics demonstrate that our learned anchors jointly satisfy trajectory recoverability and ECG--visual alignability, justifying our reformulation of sparse cardiac synthesis as a recoverability--alignability anchor discovery problem.
\end{itemize}
\section{Related Work}

\textbf{Cardiac Dynamic Synthesis from Medical Imaging}. 
Cardiac dynamics synthesis reconstructs a full cardiac cycle from partial observations, enabling spatiotemporal and controllable completion~\cite{algethami2025generative, liu2026exploiting, PUNG2026103308}. Existing CMR studies under limited acquisition address temporal interpolation~\cite{Kim_2024_CVPR, You_2025_ICCV}, motion reconstruction~\cite{bubeck2025latent, LAUMER2025103434}, super-resolution~\cite{11333374, zhou2025diffusion}, and conditional generation~\cite{liu2025sagcnet, ma2025cardiacflow}. For ECHO, video synthesis methods have been studied under diverse conditioning settings~\cite{reynaud2025echoflow, fang2026ecgflowcmr, zhou2024heartbeat, oladokun2026echolvfm, Zhang_2026_CVPR}. Despite these advances, existing methods typically rely on richer visual evidence, modality-specific assumptions, or global conditioning signals, making them less suited to one-shot cardiac synthesis. CardioFAD instead learns motion-salient keyframes that are both trajectory-sufficient and ECG-alignable, and uses them as anchors for two-stage generation across 3D+t CMR and 2D+t ECHO.

\textbf{Diffusion-based Video Generation}. 
Diffusion models have become a dominant paradigm for video generation~\cite{Rombach_2022_CVPR, Peebles_2023_ICCV, lu2024vdt, hu2025lamd, zhang2026videorepa}. Controllable generation further incorporates constraints such as reference frames, motion guidance, and temporally varying control signals~\cite{NEURIPS2023_180f6184, 10.1145/3641519.3657518, 10.1145/3680528.3687652, zhang2026frame, xing2024dynamicrafter}. Recent works show that conditioning on structured temporal information can substantially improve temporal coherence and enable targeted manipulation~\cite{Feng_2025_ICCV, Liu_2025_ICCV, Geng_2025_CVPR, Bahmani_2025_CVPR, Zhang_2025_CVPR, Zhang_2025_MotionPro}. Despite these advancements, existing conditioning does not explicitly handle the extreme observation sparsity in our task context. Our two-stage strategy first generates keyframes and then interpolates remaining frames, enabling effective guidance from sparse observations and physiological signals.

\textbf{Keyframe Selection for Video Processing}. 
Keyframe selection aims to identify a compact set of frames that best summarizes video content~\cite{10.1145/3721238.3730664, zhu2025focus, Tang_2025_CVPR, guo2025logicinframes}. It is widely used for video understanding~\cite{Tang_2025_CVPR, lee2025refocus, Liu_2025_CVPR, Cao_2025_ICCV, Sun_2025_ICCV}, and for efficient modeling by focusing computation on informative locations~\cite{Buch_2025_CVPR, zhu2025focus, wang2025videocompressa, Hu_2025_CVPR, kumar-2025-language}. For video generation, keyframes can serve as fundamental anchors that capture salient states, enabling coarse-to-fine synthesis~\cite{Bae_2025_ICCV, 10943599, wang2025keyvid, geng2024text, Bigata_2025_CVPR}. Nevertheless, existing keyframe-based methods typically assume single-modality inputs and rely on predefined or externally supervised anchors, with limited consideration of cross-modal alignment. In contrast, CardioFAD learns self-emergent keyframes from time-series guidance, selecting anchors that are both intra-modally trajectory-sufficient and inter-modally alignable for two-stage cardiac synthesis.
\section{Methodology}

\subsection{Problem Formulation}
We aim to reconstruct a complete cardiac cycle from sparse visual observations, guided by synchronized time-series (ECG). We validate our approach on two modalities: \textbf{(i) 3D+t CMR synthesis}, recovering a full 3D+t sequence from a single 2D slice or 3D volume; and \textbf{(ii) 2D+t ECHO video synthesis}, generating a complete 2D+t video from a single frame. CMR uses paired patient-specific ECG, while ECHO uses a shared patient-independent phase template for temporal guidance.


\subsection{Focus: Cardiac Dynamic Keyframe Selection}
\label{focus}
Given a cardiac sequence and an ECG signal, our visual and ECG encoders first extract trajectory tokens $\mathbf{z}_t$ and time-series embeddings $\mathbf{h}_t$, respectively. Our tokenization scheme is elaborately designed to map these heterogeneous inputs into a shared temporal length $T$ for subsequent processing.

\textbf{Keyframe selection}.
Our keyframe selector operates in the cardiac state space, which yields learnable keyframes directly from the input ECG, empowering our tokenizer to perceive clinically meaningful phases in a self-emergent manner. Let $T$ denote the sequence length in the latent space. A lightweight transformer~\cite{NIPS2017_3f5ee243} $f_\phi$ consumes the ECG tokens and outputs scalar scores $\ell_t$ as
\begin{equation}
    \ell_{1:T} = f_\phi(\mathbf{h}_{1:T}), \qquad
    q_t = \frac{\exp(\ell_t)}{\sum_{k=1}^{T} \exp(\ell_k)},
\label{q_t}
\end{equation}
where $q_t$ is the keyframe probability at time $t$.
We produce a binary selection mask $m_t \in \{0,1\}$ via straight-through Gumbel-Softmax top-$K$~\cite{jang2017categorical}. This yields a keyframe set $\mathcal{K} = \{t \mid m_t = 1\}$ of size $K$. To ensure that keyframes are informative enough, we adopt the following principle:

\noindent \textbf{Trajectory Sufficiency Principle (TSP).} \textit{The selected keyframe set should be sufficient to recover the entire cardiac trajectory.}

To enforce this, we require that the model should reconstruct the entire input sequence with minimal error, making keyframes a maximally informative and compact summary of intra-modal dynamics. Specifically, we mask out non-keyframe latents by setting $\tilde{\mathbf{z}}_t = m_t \mathbf{z}_t$. A 3D UNet $G_\psi$ then attempts to reconstruct the complete latent sequence from this sparse input as $\hat{\mathbf{z}}_{1:T} = G_\psi(\tilde{\mathbf{z}}_{1:T})$. The reconstruction loss is computed exclusively on the non-keyframe steps:
\begin{equation} 
\mathcal{L}_{focus} = \frac{1}{T - K} \sum_{t \notin \mathcal{K}} \|\hat{\mathbf{z}}_t - \mathbf{z}_t\|_1.
\end{equation}
This principle aligns with the intuition that the learned keyframes should encapsulate the most informative and predictive dynamics of the visual modality.



\subsection{Align: Cross-Modal Cardiac Trajectory Alignment}
\label{align}
To bridge the representation gap and mitigate phase misalignment, we synchronize the visual trajectory with the ECG signals, explicitly transforming the time-series condition into reliable guidance.

\textbf{Cross-modal contrastive learning.}
We establish a temporally aware alignment matrix $A$ to encode correspondences between the modalities based on cardiac phases. With both modalities mapped to a shared temporal length 
$T$, $d_{\mathrm{time}}(k,t)$ denotes the circular distance between normalized cardiac phases, accounting for cycle wrap-around:
\begin{equation}
    A_{k,t} 
    = \frac{\exp\big(- d_{\mathrm{time}}(k,t)^2 / (2\sigma^2)\big)}
           {\sum_{i,j} \exp\big(- d_{\mathrm{time}}(i,j)^2 / (2\sigma^2)\big)},
\label{A_matrix}
\end{equation}
where $\sigma$ controls the temporal neighborhood width.
Using $A$ as a soft alignment target, we compute a cross-modal contrastive loss. For the ECG-to-Visual direction, the target distribution $p$ and the predicted similarity distribution $\pi$ are formulated as:
\begin{equation}
    p^{\mathrm{E}\rightarrow\mathrm{V}}_{t,k} 
    = \frac{A_{k,t}}{\sum_j A_{j,t}},
    \qquad
    \pi^{\mathrm{E}\rightarrow\mathrm{V}}_{t,k} 
    = \frac{\exp(S_{k,t} / \tau)}{\sum_j \exp(S_{j,t} / \tau)},
\label{distribution_p}
\end{equation}
where $S_{k,t}$ is the cosine similarity between the mean-pooled ECG and visual tokens, and $\tau$ is a temperature parameter. The alignment loss for this direction is defined as:
\begin{equation}
    \mathcal{L}_{\mathrm{align}}^{\mathrm{E}\rightarrow\mathrm{V}}
    = - \frac{1}{T} \sum_{t=1}^{T} \sum_{k=1}^{T} 
      p^{\mathrm{E}\rightarrow\mathrm{V}}_{t,k} 
      \log \pi^{\mathrm{E}\rightarrow\mathrm{V}}_{t,k}.
\end{equation}
Symmetrically, we define a visual-to-ECG loss $\mathcal{L}_{align}^{V \rightarrow E}$ by normalizing $A$ along the other axis, yielding the base cross-modal alignment objective as $\mathcal{L}_{\mathrm{align}} 
    = \frac{1}{2} \Big(
        \mathcal{L}_{\mathrm{align}}^{\mathrm{E}\rightarrow\mathrm{V}}
        + 
        \mathcal{L}_{\mathrm{align}}^{\mathrm{V}\rightarrow\mathrm{E}}
      \Big).$

\textbf{Keyframe-weighted alignment.} In the context of time-series-guided generation, we argue that keyframes play the most significant role in explaining the cross-modal relationship. To this end, we introduce the following principle:

\noindent\textbf{Anchor Alignment Principle (AAP).}
\textit{Keyframes serve as cross-modal anchors that dominate alignment supervision.}

To implement AAP, we assign higher alignment responsibility to keyframes by introducing a symmetric keyframe importance term based on the probabilities $q$ in Eqn.~\ref{q_t}, yielding $W_{k,t} = \frac{q_k + q_t}{\sum_{i,j} (q_i + q_j)}$. We then incorporate this term to reformulate the base alignment matrix $A_{k,t}$ (Eqn.~\ref{A_matrix}) into:
\begin{equation} 
\tilde{A}_{k,t} = \frac{\exp\big(- d_{\mathrm{time}}(k,t)^2 / (2\sigma^2)\big)\, W_{k,t}} {\sum_{i,j} \exp\big(- d_{\mathrm{time}}(i,j)^2 / (2\sigma^2)\big)\, W_{i,j}}. 
\end{equation}
Because $W_{k,t}$ scales with the keyframe probabilities, the dynamically salient keyframes automatically receive higher weights in $\tilde{A}$. This ensures that temporally synchronized and keyframe-like moments contribute more to the alignment. We replace the base matrix $A_{k,t}$ with the keyframe-weighted matrix $\tilde{A}_{k,t}$ in Eqn.~\ref{distribution_p} to compute the target distribution $p$, seamlessly enforcing the AAP in the overall loss $\mathcal{L}_{align}$. Ultimately, the overall loss of the tokenizer is the weighted sum of the reconstruction loss and the two principle-guided auxiliary objectives:
\begin{equation}
\mathcal{L}_{tok} = \mathcal{L}_{\mathrm{rec}} + \lambda_1 \mathcal{L}_{\mathrm{focus}} + \lambda_2 \mathcal{L}_{\mathrm{align}}.
\label{loss}
\end{equation}

\subsection{Diffuse: Two-Stage Keyframe-to-Sequence Generation}
\label{diffuse}



\textbf{Conditional video DiT}.
We adopt a modified video DiT~\cite{ma2025latte, lu2024vdt} as our generative backbone. It comprises a stack of alternating spatial and temporal transformer blocks. More specifically, the spatial blocks attend to spatiotemporal patches and receive global conditioning from the sparse visual observation's latent via AdaLN~\cite{Peebles_2023_ICCV}, while temporal blocks attend along the cardiac phase axis, receiving per-frame conditioning from the encoded ECG latents $\mathbf{h}_t$, the observed frame index, and the keyframe probabilities $q_t$ via cross-attention. 

\textbf{Unified masked diffusion}.
Rather than using separate models, we formulate both keyframe generation and sequence interpolation as a unified masked diffusion task in the latent space. Given the keyframe mask $m\in\{0,1\}^T$ predicted by \textbf{Focus} (Sec.~\ref{focus}), we randomly sample a training mode at each iteration: (A) \emph{Keyframe generation} uses $m$ to activate the diffusion loss on keyframe positions, while (B) \emph{Frame interpolation} uses $1-m$ to activate the loss on non-keyframe positions. During training, we sample mode $s\in\{\mathrm{A},\mathrm{B}\}$ with probability $p(s)$ and optimize a weighted objective $\mathcal{L}_{diff} = \lambda_s \mathcal{L}_s$ to balance anchor learning and trajectory completion. During inference, we first sample keyframe latents (mode A) and subsequently generate non-keyframes conditioned on these anchors (mode B). Further details are provided in Appendix~\ref{app:generation}.




\section{Experiments}

\begin{table}[t]
\centering
\setlength{\tabcolsep}{2.5mm}
\renewcommand\arraystretch{1.0}
\caption{\textbf{Quantitative results for 3D+t CMR synthesis on UKBB dataset}.}
\resizebox{1.0\linewidth}{!}{
\begin{tabular}{lccccc|ccccc}
\toprule
\multirow{2}{*}{\textbf{Methods}} & \multicolumn{5}{c|}{\textbf{Volume2Seq}} & \multicolumn{5}{c}{\textbf{Slice2Seq}} \\
\cmidrule(lr){2-6}\cmidrule(lr){7-11}
  & PSNR$\uparrow$ & SSIM$\uparrow$ & LPIPS$\downarrow$ & FID$\downarrow$ & FVD$\downarrow$ & PSNR$\uparrow$ & SSIM$\uparrow$ & LPIPS$\downarrow$ & FID$\downarrow$ & FVD$\downarrow$\\
\midrule
VideoComposer~\cite{NEURIPS2023_180f6184}   & 28.15 & 0.843 & 0.107 & 128.76 & 1762.34 & 21.86 & 0.497 & 0.289 & 314.85 & 4574.61 \\
MotionCtrl~\cite{10.1145/3641519.3657518}   & 28.46 & 0.825 & 0.104 & 119.52 & 1334.87 & 21.25 & 0.488 & 0.294 & 365.97 & 4914.27 \\
Tora~\cite{Zhang_2025_CVPR}                 & 29.74 & 0.854 & 0.092 & 97.30 & 1025.57 & 23.10 & 0.539 & 0.251 & 267.92 & 3109.75 \\
MotionPro~\cite{Zhang_2025_MotionPro}       & 27.88 & 0.817 & 0.116 & 145.39 & 1887.96 & 22.09 & 0.501 & 0.269 & 292.17 & 3756.61 \\
\midrule
\rowcolor{tableblue!80}
CardioFAD (One-stage)     &  30.51  &  0.861 &  0.082 &  82.48 &  1007.63 &
 23.75 &  0.559 &  0.231 &  211.49 &  2785.30 \\
\rowcolor{tablepurple!70}
\textbf{CardioFAD (Ours)}     &  \textbf{32.41}  &  \textbf{0.885} &  \textbf{0.064} &  \textbf{59.96} &  \textbf{736.01} &
 \textbf{25.60} &  \textbf{0.587} &  \textbf{0.198} &  \textbf{163.12} &  \textbf{1847.73} \\
\bottomrule
\end{tabular}
}
\vspace{-12pt}
\label{tab:CMR}
\end{table}

\begin{table}[t]
\centering
\setlength{\tabcolsep}{1.0mm}
\renewcommand\arraystretch{1.0}
\caption{\textbf{Quantitative results for 2D+t ECHO synthesis on three datasets}.}
\resizebox{\linewidth}{!}{
\begin{tabular}{lccccc|ccccc|ccccc}
\toprule
\multirow{2}{*}{\textbf{Methods}} & \multicolumn{5}{c|}{\textbf{EchoNet-Dynamic}} & \multicolumn{5}{c|}{\textbf{EchoNet-Pediatric (A4C)}} & \multicolumn{5}{c}{\textbf{EchoNet-Pediatric (PSAX)}} \\
\cmidrule(lr){2-6}\cmidrule(lr){7-11}\cmidrule(lr){12-16}
  & PSNR$\uparrow$ & SSIM$\uparrow$ & LPIPS$\downarrow$ & FID$\downarrow$ & FVD$\downarrow$ & PSNR$\uparrow$ & SSIM$\uparrow$ & LPIPS$\downarrow$ & FID$\downarrow$ & FVD$\downarrow$ & PSNR$\uparrow$ & SSIM$\uparrow$ & LPIPS$\downarrow$ & FID$\downarrow$ & FVD$\downarrow$\\
\midrule
\rowcolor{tablegray}
\multicolumn{16}{c}{\textit{General Video Generation}} \\
VideoComposer~\cite{NEURIPS2023_180f6184}     & 22.78 & 0.539 & 0.109 & 48.25 & 284.16 & 27.49 & 0.633 & 0.089 & 34.74 & 346.31 & 27.26 & 0.621 & 0.084 & 42.19 & 389.07 \\
Tora~\cite{Zhang_2025_CVPR}   & 23.90 & 0.548 & 0.102 & 52.06 & 241.96 & 27.86 & 0.638 & 0.082 & 28.94 & 361.09 & 28.37 & 0.638 & 0.081 & 35.52 & 347.21 \\
\midrule
\rowcolor{tablegray}
\multicolumn{16}{c}{\textit{ECHO Video Generation}} \\
HeartBeat~\cite{zhou2024heartbeat}   & 21.24 & 0.527 & 0.122 & 58.97 & 376.51 & 26.71 & 0.629 & 0.088 & 37.56 & 415.71 & 27.65 & 0.630 & 0.079 & 45.02 & 407.26 \\
EchoNet-Synthetic~\cite{reynaud2024echonet}     & 23.58 & 0.540 & 0.107 & 49.87 & 267.35 & 26.08 & 0.624 & 0.092 & 39.93 & 408.12 & 26.23 & 0.612 & 0.090 & 48.82 & 432.50 \\
ECHOPulse~\cite{li2025echopulse}     &  25.63 &  0.568 &  0.081 &  37.05 &  196.28 & 28.14 & 0.641 & 0.080 & 29.63 & 357.74 & 28.50 & 0.639 & 0.080 &  37.57 &  367.46 \\
\midrule
\rowcolor{tableblue!80}
CardioFAD (One-stage)  &  25.14 & 0.556 & 0.089 & 46.82 & 232.17 &
 28.32 &  0.643 &  0.078 &  29.23 &  339.15 
&  28.77 &  0.640 &  0.079 & 39.17 &  381.06 \\
\rowcolor{tablepurple!70}
\textbf{CardioFAD (Ours)}  &  \textbf{27.73}  &  \textbf{0.596} &  \textbf{0.067} &  \textbf{26.92} &  \textbf{137.70} &
 \textbf{29.48} &  \textbf{0.657} &  \textbf{0.073} &  \textbf{23.21} &  \textbf{286.29} 
&  \textbf{29.59} &  \textbf{0.645} &  \textbf{0.071} &  \textbf{27.12} &  \textbf{293.86} \\
\bottomrule
\end{tabular}
}
\vspace{-12pt}
\label{tab:ECHO}
\end{table}

\subsection{Experimental Settings}
\label{sec:exp_setting}

\textbf{Datasets.}
\textbf{(i) 3D+t CMR synthesis.} We use the UK Biobank (UKBB)~\cite{PETERSEN20168} dataset with paired short-axis CMR sequences and ECG recordings, which contains 4,379 samples with 50 frames per subject. \textbf{(ii) ECHO video synthesis.} We evaluate on EchoNet-Dynamic~\cite{ouyang2020video} and EchoNet-Pediatric~\cite{REDDY2023482}. For the latter, we treat A4C and PSAX as two separate datasets. For ECHO datasets without ECG, we use a patient-agnostic canonical phase template as a weak pseudo-ECG surrogate (see Appendix~\ref{app:pseudo_ecg}). This template is shared across samples and does not use test-time ED/ES annotations, contours, volume tracings, or any patient-specific temporal labels. Perturbation results in Appendix~\ref{app:perturbation} further show that CardioFAD mainly relies on coarse phase cues rather than precise waveform details. 


\textbf{Implementation Details}.
We adopt LeanVAE~\cite{Cheng_2025_ICCV} as the visual tokenizer and a lightweight ViT~\cite{dosovitskiy2020image} as the ECG encoder. We set the number of keyframes to $K=10/8$. The LeanVAE is trained jointly with the keyframe selector and alignment modules, which is justified in Appendix~\ref{app:frozen_tok}. For generation, we adopt rectified flow~\cite{lipman2023flow, liu2023flow} as the training objective. We set the mode sampling probabilities to $p(A) = p(B) = 0.5$, and the objective weights to $\lambda_A = 1.5$ and $\lambda_B = 0.5$. Please refer to Appendix~\ref{app:imple} for more training details and hyperparameter settings.

\textbf{Evaluation Metrics}. 
We use five metrics for comparisons: Peak Signal-to-Noise Ratio (PSNR), Structural Similarity Index (SSIM), Learned Perceptual Image Patch Similarity (LPIPS), Fréchet Inception Distance (FID)~\cite{NIPS2017_8a1d6947} for visual quality, and Fréchet Video Distance (FVD)~\cite{unterthiner2018towards} for temporal consistency. Beyond these generative metrics, we assess the cardiac functional consistency of synthesized ECHO videos via ejection fraction (EF) estimation. Following~\cite{reynaud2025echoflow, reynaud2024echonet}, we train an EF regressor on training videos, and then apply it to generated videos to obtain predicted EF values (details in Appendix~\ref{app:EF}). We report $R^2$, MAE, and RMSE against ground-truth EF. 

\textbf{Comparison Methods}.
Since no prior method matches our extreme-sparsity CMR setting, we compare with representative controllable video generation baselines~\cite{NEURIPS2023_180f6184, 10.1145/3641519.3657518, Zhang_2025_CVPR, Zhang_2025_MotionPro}, and include ECHO-specific baselines~\cite{reynaud2024echonet, li2025echopulse, zhou2024heartbeat} for ECHO synthesis. All baselines are re-trained under the same data preprocessing, observation protocol, and matched training setups. To better isolate the effect of our design, we also report a strong internal one-stage baseline, \textit{\textbf{CardioFAD (One-stage)}}, that shares the same tokenizer, backbone, and training setup as CardioFAD, but removes the proposed keyframe anchoring and two-stage generation. This controlled baseline provides the \textbf{fairest comparison} for attributing gains to the Focus--Align--Diffuse design rather than architecture or training differences.

\begin{figure*}[t]
\centering
\includegraphics[width=\linewidth]{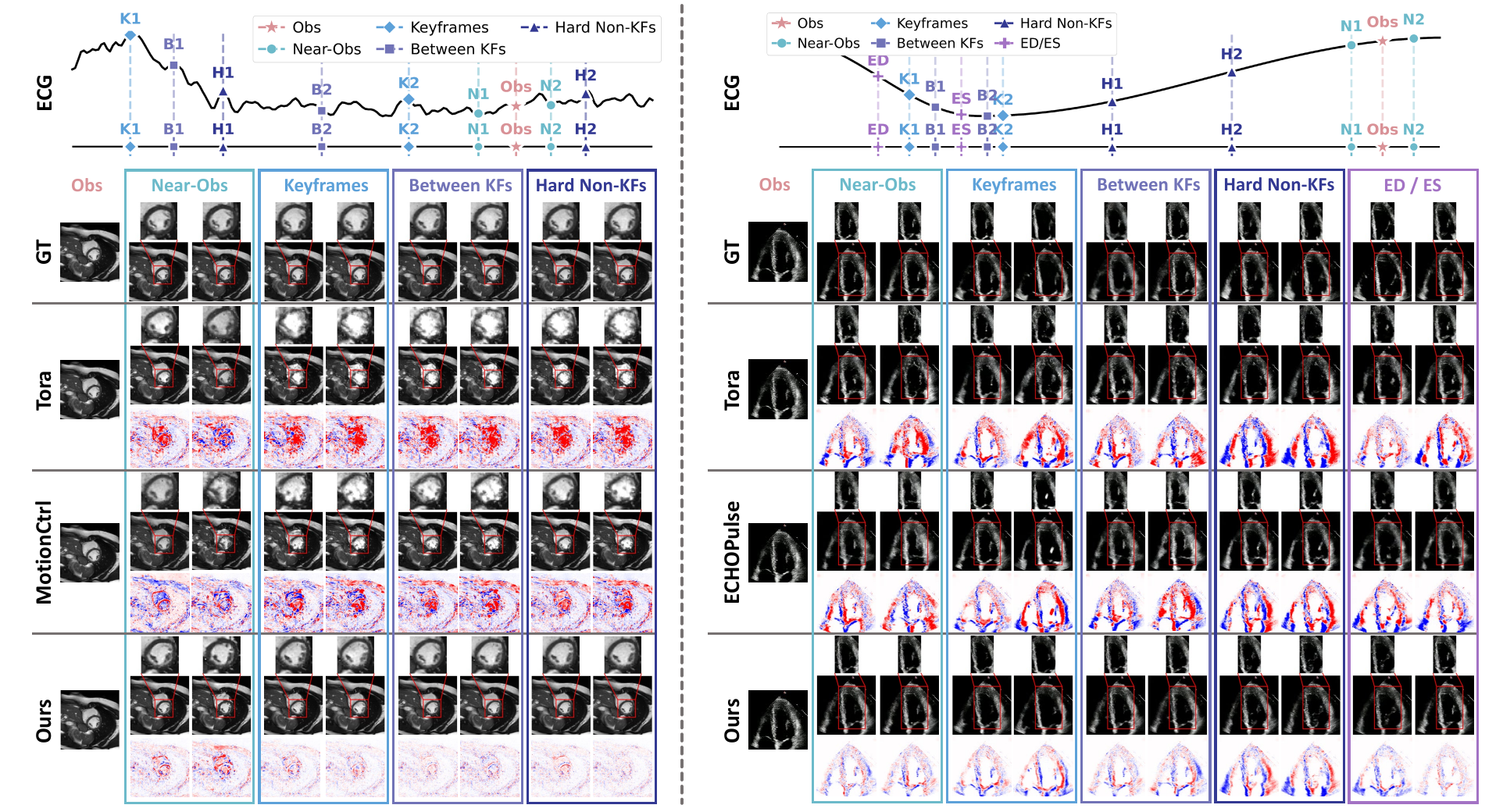}
\vspace{-14pt}
\caption{\textbf{Qualitative comparisons on CMR (left) and ECHO (right)}. Frames are grouped by temporal roles, with top ECG curves providing physiological context. Zoomed-in crops and difference heatmaps are provided. Compared to baselines, CardioFAD achieves higher visual fidelity and temporal consistency, avoiding phase drift particularly at challenging positions.}
\vspace{-6pt}
\label{qualitative}
\end{figure*}

\begin{table*}[t]
\centering
\begin{minipage}[t]{0.48\textwidth}
  \centering
  \setlength{\tabcolsep}{2.0mm}
  \renewcommand\arraystretch{0.9}
  \caption{\textbf{Cardiac functional consistency} evaluated by EF estimation on EchoNet-Dynamic.}
  \resizebox{0.90\linewidth}{!}{
    \begin{tabular}{lccc}
      \toprule
      {\textbf{Methods}} 
      & $R^2$$\uparrow$ & MAE$\downarrow$ & RMSE$\downarrow$ \\
      \midrule
      \rowcolor{tablegray}
      \multicolumn{4}{c}{\textit{General Video Generation}} \\
      VideoComposer~\cite{NEURIPS2023_180f6184}   & 0.57 & 6.03 & 8.49   \\
      Tora~\cite{Zhang_2025_CVPR}     & 0.55 & 6.10 & 8.76  \\
      \midrule
      \rowcolor{tablegray}
      \multicolumn{4}{c}{\textit{ECHO Video Generation}} \\
      HeartBeat~\cite{zhou2024heartbeat}   & 0.57 & 6.08 & 8.53  \\
      EchoNet-Synthetic~\cite{reynaud2024echonet}     & 0.71 & 5.07 & 7.09  \\
      ECHOPulse~\cite{li2025echopulse}     & 0.78 & 4.62 & 6.08  \\
      \midrule
      \rowcolor{tableblue!80}
      CardioFAD (One-stage)    &  0.69  &  4.98 &  7.19   \\
      \rowcolor{tablepurple!70}
      \textbf{CardioFAD (Ours)}       &  \textbf{0.82}  &  \textbf{4.17} &  \textbf{5.59}   \\
      \bottomrule
    \end{tabular}
  }
  \label{tab:ECHO_clinical}
  \vspace{-12pt}
\end{minipage}\hfill
\begin{minipage}[t]{0.48\textwidth}
  \centering
  \setlength{\tabcolsep}{0.5mm}
  \renewcommand\arraystretch{1.0}
  \caption{\textbf{Principle-guided keyframe learning ablation} on EchoNet-Dynamic.}
  \resizebox{\linewidth}{!}{
    \begin{tabular}{lccccc}
      \toprule
      \textbf{Methods} & PSNR$\uparrow$ & SSIM$\uparrow$ & LPIPS$\downarrow$ & FID$\downarrow$ & FVD$\downarrow$ \\
      \midrule
      \rowcolor{tablegray}
      \multicolumn{6}{c}{\textit{Keyframe Learning Principle}} \\
      w/o TSP              & 25.98 & 0.570 & 0.082 & 41.83 & 203.37   \\
      w/o AAP              & 26.74 & 0.581 & 0.079 & 33.96 & 180.29  \\
      w/o TSP \& w/o AAP              & 25.58 & 0.566 & 0.085 & 44.65 & 239.81   \\
      \midrule
      \rowcolor{tablegray}
      \multicolumn{6}{c}{\textit{Keyframe Selection Strategy}} \\
      w/ uniform keyframes       & 25.94 & 0.572 & 0.081 & 40.41 & 188.36 \\
      w/ random keyframes        & 25.17 & 0.554 & 0.091 & 45.59 & 251.03 \\
      w/ GT ED/ES                & 26.20 & 0.572 & 0.083 & 42.76 & 202.41 \\
      w/ GT ED/ES + Supplement   & 26.83 & 0.587 & 0.074 & 34.53 & 184.97 \\
      \rowcolor{tablepurple!70}
      \textbf{CardioFAD}        &  \textbf{27.73}  &  \textbf{0.596} &  \textbf{0.067} &  \textbf{26.92} &  \textbf{137.70}  \\
      \bottomrule
    \end{tabular}
    } 
  \label{tab:ablation_principles}
  \vspace{-12pt}
\end{minipage}
\end{table*}

\subsection{Comparison with State-of-the-Art}

We consider three \emph{extreme-sparsity} settings: \textbf{Volume2Seq} and \textbf{Slice2Seq} for 3D+t CMR, and \textbf{Frame2Video} for 2D+t ECHO, each reconstructing the full sequence from a single observed volume/slice/frame. One-shot observation corresponds to $B{=}1$, obtained by randomly sampling one visual observation, where $B$ denotes the number of available visual observations. 


\noindent\textbf{Quantitative Results}.
As shown in Tables~\ref{tab:CMR} and~\ref{tab:ECHO}, the internal baseline \textit{CardioFAD (One-stage)} is already competitive with baselines, and surpasses them in multiple settings, validating our strong architectural foundation. Building upon this, \textit{CardioFAD (Ours)} yields further improvements, particularly in temporal consistency. This substantial gain directly confirms our core motivation that explicitly anchoring generation with learned keyframes via a two-stage process is essential for resolving temporal ambiguity and ensuring coherent dynamics under extreme sparsity.

\noindent\textbf{Cardiac Functional Consistency}.
As reported in Table~\ref{tab:ECHO_clinical}, CardioFAD achieves the best EF regression performance on EchoNet-Dynamic among all baselines, indicating improved preservation of EF-related functional dynamics in the synthesized videos.

\noindent\textbf{Qualitative Results}.
In Fig.~\ref{qualitative}, instead of uniformly sampling frames, we uniquely organize frames by their temporal roles alongside ECG context. While baselines suffer from motion inconsistency and blurring, especially at intermediate and hard positions, CardioFAD maintains sharp, phase-consistent dynamics across the entire cardiac cycle. This visually validates that our learned motion-salient keyframes act as robust anchors, successfully regularizing the challenging non-keyframe generation.

\subsection{Robustness under Extreme Sparsity}
\begin{figure}[t]
    \centering
      \includegraphics[width=\linewidth]{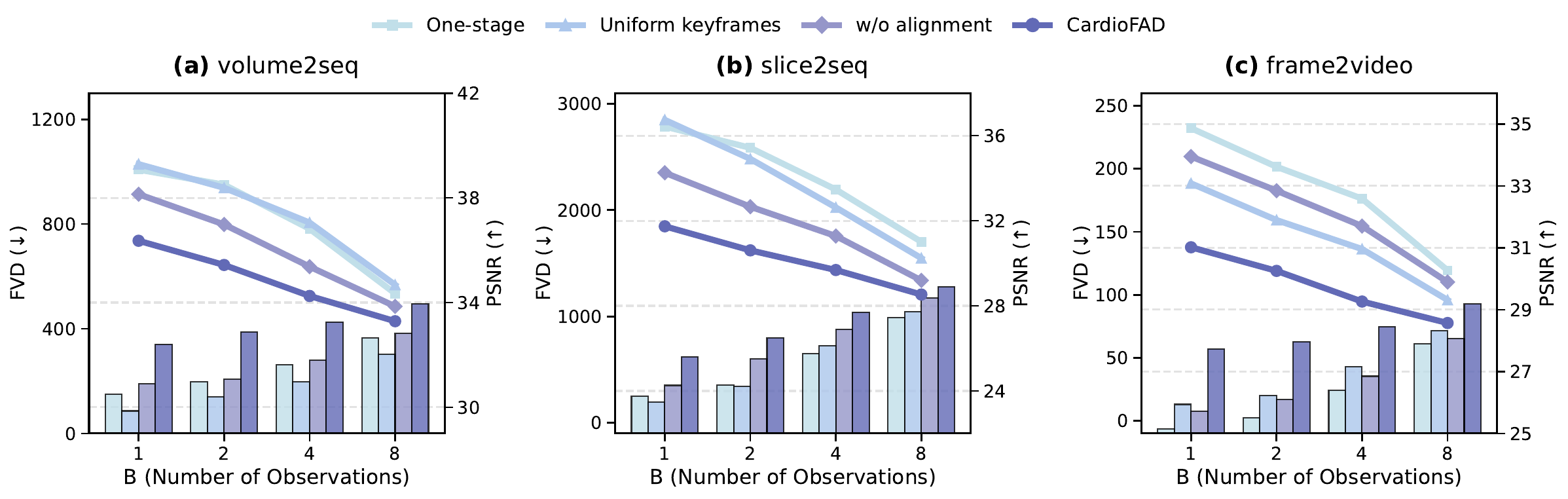}
    \vspace{-12pt}
    \caption{\textbf{Sparse observation budget}. Line charts report FVD, and bar charts report PSNR. When $B=1$ (extremely sparse), CardioFAD exhibits its greatest advantage. }
    \label{sparsity}
    \vspace{-2pt}
\end{figure}

We further evaluate $B\in\{1,2,4,8\}$ to assess robustness across observation budgets. In Fig.~\ref{sparsity}, CardioFAD achieves its largest gains at $B{=}1$, where visual evidence is most limited and temporal ambiguity is strongest. Although the gap narrows as more observations become available, CardioFAD remains competitive across all budgets, indicating robust behavior beyond the one-shot setting.

\begin{figure}[t]
    \centering
      \includegraphics[width=\linewidth]{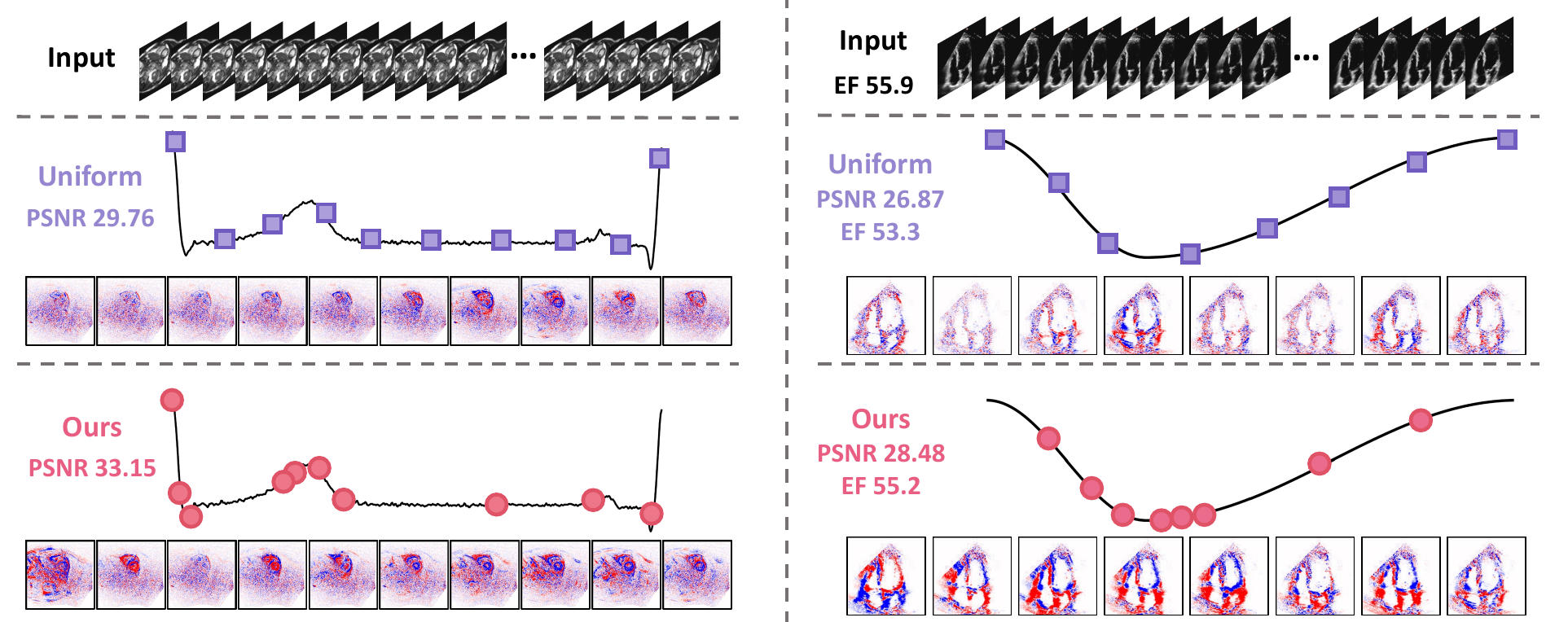}
    \vspace{-8pt}
    \caption{\textbf{Visualization of keyframe selection strategies}. Representative CMR (left) and ECHO (right) sequences with ECG overlaid. Inter-frame difference heatmaps are provided to illustrate motion magnitude. Our learned keyframes explicitly target motion-salient phases.}
    \label{keyframe_selection}
    \vspace{-10pt}
\end{figure}

\subsection{Ablation Studies}
Table~\ref{tab:ablation_principles} isolates the role of principle-guided anchor discovery. Removing either TSP or AAP degrades performance, and replacing our learned anchors with random, uniform, or GT ED/ES anchors, even when ED/ES is padded to the same budget $K=8$, consistently performs worse. These findings indicate that the gain does not come from two-stage generation alone, but from \textbf{discovering self-emergent anchors that lie in the recoverable-and-alignable paradigm}.

\textbf{Keyframe Visualization.}
Fig.~\ref{keyframe_selection} further illustrates the advantage of learned keyframes over uniform sampling. 
Inter-frame difference heatmaps indicate that our keyframes concentrate on motion-salient phases with larger anatomical changes, providing more informative anchors for two-stage generation.

Further fine-grained ablation results and analyses of robustness to ECG perturbations are provided in Appendix~\ref{app:ablation} and~\ref{app:perturbation}, respectively. These additional ablations further separate the effects of ECG guidance, cross-modal alignment, keyframe probabilities, and reference-frame indexing, highlighting that the gains do not stem from a single conditioning source.

\subsection{Principle-Targeted Diagnostics for Anchor Discovery}
Beyond standard ablations, we use principle-targeted diagnostics to test the central insight of our reformulation, i.e., effective sparse cardiac synthesis requires anchors that are not only trajectory-recoverable, but also ECG--visually alignable.

\begin{figure}[t]
    \centering
    \includegraphics[width=0.99\linewidth]{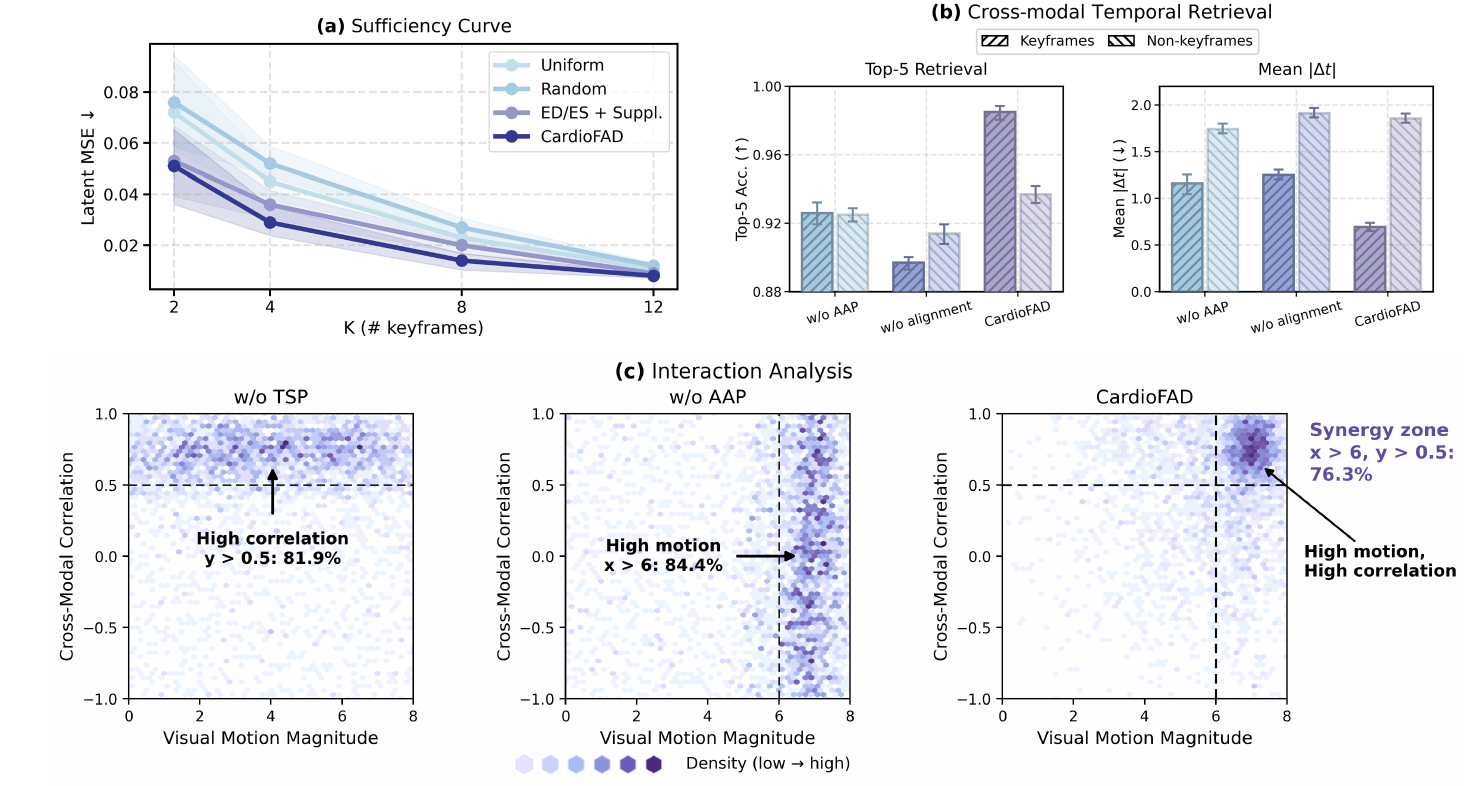}
    \vspace{-4pt}
      \caption{\textbf{(a) TSP Effectiveness:} CardioFAD achieves lower error with fewer frames, indicating higher information density. \textbf{(b) AAP Effectiveness:} Cross-modal retrieval metrics show that AAP successfully concentrates alignment capability on keyframes compared to non-keyframes. \textbf{(c) Interaction Analysis:} Visualizing keyframes in Motion-Correlation space. Removing TSP biases selection towards high-correlation frames (top), while removing AAP leads to high-motion frames (right). CardioFAD (top-right) effectively balances dynamic salience with physiological consistency.}
    \label{principle}
    \vspace{-6pt}
\end{figure}


\textbf{TSP: Maximizing Information Density}. \textit{Does CardioFAD select the most informative frames?} In Fig.~\ref{principle}(a), we measure the reconstruction error of a trajectory. We adopt a selection-agnostic Keyframe-to-Trajectory probe trained with randomly masked keyframes. CardioFAD consistently yields the lowest latent MSE, especially at low budgets ($K=2,4$). This confirms that our selector prioritizes trajectory-sufficient states that maximally preserve intra-modal motion.

\textbf{AAP: Aligning Cardiac Trajectories}.
\textit{Does AAP concentrate ECG--visual alignment on anchors?} Fig.~\ref{principle}(b) reports ECG-to-visual time-step retrieval performance using the learned embeddings, measured by Top-5 accuracy and mean absolute temporal error, $\lvert \Delta {\text{t}} \rvert$. Our model achieves satisfactory alignment accuracy specifically on keyframes compared to non-keyframes. Removing AAP degrades this anchor-specific performance, verifying that AAP concentrates alignment supervision on an anchor subset, turning ECG into reliable constraints exactly where motion is most informative.

\textbf{Interaction: Validating Recoverability--Alignability Anchors}. 
\textit{Do TSP and AAP jointly discover anchors in the recoverable-and-alignable paradigm?} Fig.~\ref{principle}(c) visualizes learned keyframes in the Motion Magnitude vs. Cross-Modal Correlation space, where motion magnitude measures adjacent latent displacement and cross-modal correlation measures visual--ECG cosine similarity. Effective anchors should exhibit both high motion salience and strong physiological correspondence. As shown, without TSP, selection is biased toward visually less informative but highly correlated frames; without AAP, it drifts toward high-motion frames with weaker ECG--visual correspondence. In contrast, CardioFAD concentrates keyframes in the top-right synergy zone, providing evidence that the \textbf{proposed reformulation discovers anchors that are both trajectory-recoverable and ECG--visually alignable}.


\begin{figure}[t]
    \centering
    \includegraphics[width=1.00\linewidth]{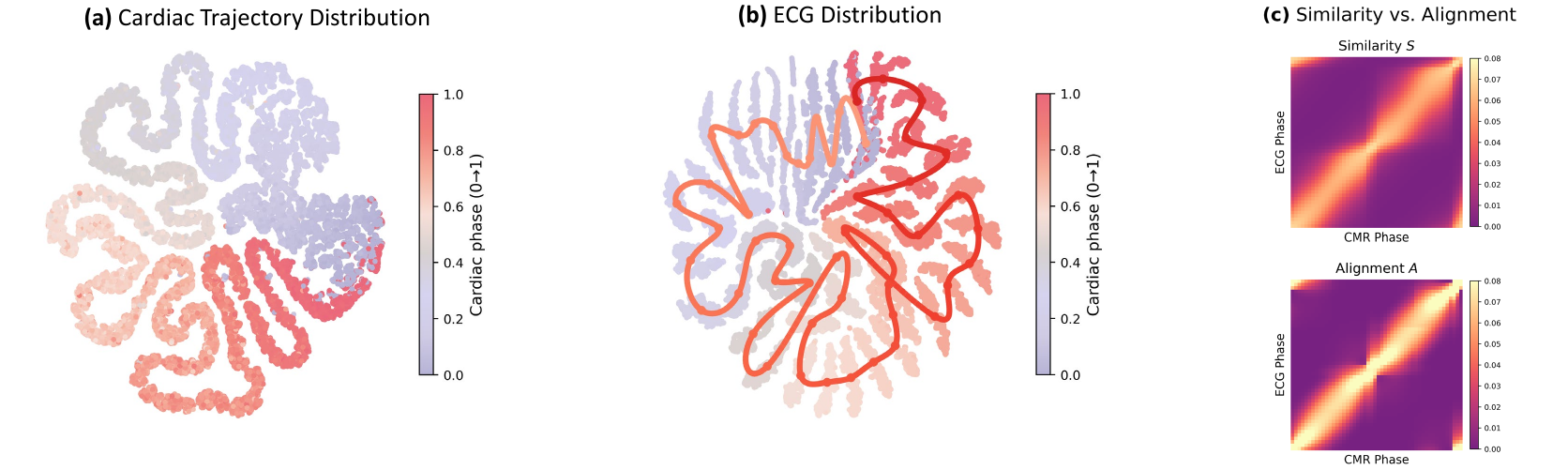}
    \vspace{-6pt}
    \caption{\textbf{Cardiac trajectory visualization}. (a) t-SNE of CMR latents. (b) t-SNE of ECG latents with a trajectory through the temporal centroids. (c) Dataset-level mean ECG-CMR similarity matrix $S$ (top) and the corresponding alignment matrix $\tilde{A}$ (bottom).}
    \label{alignment_vis}
    \vspace{-4pt}
\end{figure}

\begin{figure}[t]
    \centering
    \includegraphics[width=1.00\linewidth]{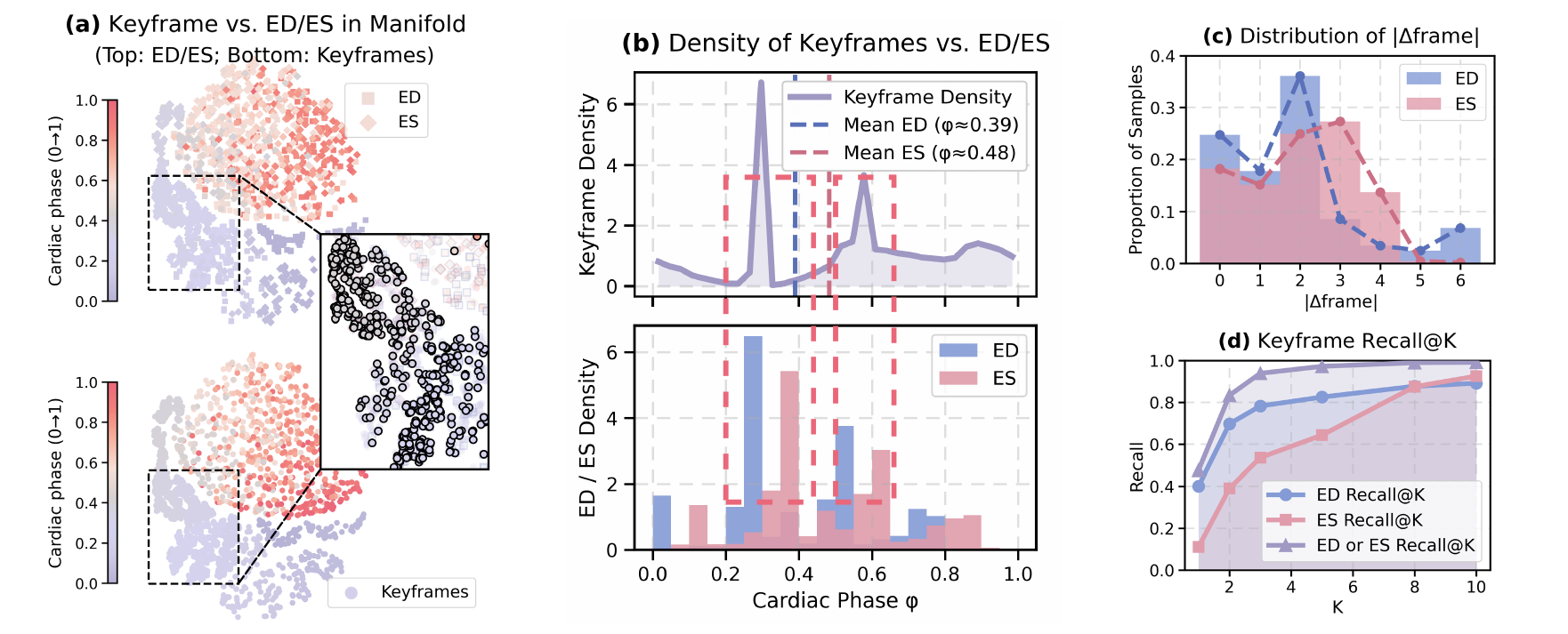}
    \vspace{-6pt}
      \caption{\textbf{Keyframe distribution visualization}. (a) GT ED/ES frames (top) and our selected keyframes (bottom) on the cardiac manifold. ED/ES-matching keyframes are outlined in black. (b) Mean keyframe phase density and ED/ES phase histograms (dashed lines: mean phases). (c) Distribution of $\lvert \Delta {\text{frame}} \rvert$ to the nearest keyframe from each ED/ES frame. (d) Keyframe Recall@K.}
    \label{keyframe_vis}
    \vspace{-10pt}
\end{figure}

\subsection{Visualization and Analysis}
\noindent\textbf{Cardiac Trajectory Alignment}.
t-SNE visualizations show that both CMR (Fig.~\ref{alignment_vis}(a)) and ECG (Fig.~\ref{alignment_vis}(b)) latents form unified, smooth manifolds with consistent phase ordering. These trends support the promotion of phase-aware cross-modal organization across modalities. The alignment matrix $\tilde{A}$ (Fig.~\ref{alignment_vis}(c)) exhibits sharper diagonal concentration than raw cosine similarity $S$, revealing that CardioFAD establishes phase-aware synchronization between the two modalities.



\noindent\textbf{Keyframe Distribution}.
We further validate the physiological relevance of our self-discovered keyframes against GT ED/ES annotations on ECHO datasets in Fig.~\ref{keyframe_vis}. \textit{Qualitatively}, our keyframes strongly co-localize with ED/ES in the latent manifold (Fig.~\ref{keyframe_vis}(a)) and exhibit sharp probability peaks near clinical phases (Fig.~\ref{keyframe_vis}(b)), demonstrating unsupervised alignment with motion-salient landmarks. \textit{Quantitatively}, temporal proximity is impressive, with over 85\% of GT ED/ES frames lying within a $\pm3$ frame window of a selected keyframe (Fig.~\ref{keyframe_vis}(c)), i.e., ED/ES are typically only a few frames from at least one learned keyframe. Consequently, Recall@K rises rapidly to 99.1\% at $K=8$ (Fig.~\ref{keyframe_vis}(d)), confirming that our method effectively captures discriminative cardiac events.
\section{Conclusion}
\label{sec:conclusion}

In this work, we introduce CardioFAD, a time-series-aware diffusion framework that reformulates sparse cardiac dynamic synthesis as a recoverability--alignability anchor discovery problem. Guided by TSP and AAP, CardioFAD learns self-emergent motion-salient anchors that are sufficient for visual trajectory recovery and reliable for ECG--visual alignment. These anchors enable keyframe-to-sequence generation that reduces temporal ambiguity, bridges visual anatomy with physiological signals, and achieves strong visual and functional preservation under one-shot scenarios.

\textbf{Limitations.} Despite the gains in Slice2Seq performance, inferring motion from 2D slices remains ambiguous and challenging. Assessing the clinical fidelity of generated CMR without explicit annotations such as segmentation labels or ED/ES phases is an open challenge we aim to address next. In addition, our current validation focuses on periodic cardiac motion, and extending the framework to non-periodic dynamics or general video generation remains to be explored.

\section*{Acknowledgments}
The computations described in this research were performed using the Baskerville Tier 2 HPC service. Baskerville was funded by the EPSRC and UKRI through the World Class Labs scheme (EP/T022221/1) and the Digital Research Infrastructure programme (EP/W032244/1) and is operated by Advanced Research Computing at the University of Birmingham.





\bibliographystyle{abbrv} 
\bibliography{neurips_2026}


\clearpage
\appendix


\section*{Appendix}

\section{Datasets}
\label{app:dataset}
The details of the CMR and ECHO datasets used in our experiments are provided in Table~\ref{tab:dataset_details}.

\noindent\textbf{(i) 3D+t CMR dataset.} We use the UK Biobank (UKBB)~\cite{PETERSEN20168} dataset, which contains 4,379 paired short-axis CMR sequences and ECG recordings, with 50 frames per subject. We use 3,503 subjects for training and 438 each for validation and testing. All CMR volumes are resized to $128\times128\times13$ and temporally aligned to a single cardiac cycle. For each subject, resting 12-lead ECG signals were parsed from the XML-encoded UK Biobank records and converted to millivolts using acquisition metadata. All leads were synchronously sampled at 500\,Hz. R-peaks were detected using NeuroKit2, and a representative R--R interval was selected to define a single-cycle electrical description. 

\noindent\textbf{(ii) ECHO datasets.} We use two public echocardiography datasets, EchoNet-Dynamic~\cite{ouyang2020video} and EchoNet-Pediatric~\cite{REDDY2023482}, for evaluation. EchoNet-Dynamic contains only A4C views, and we adopt its official train/validation/test split. EchoNet-Pediatric contains both A4C and PSAX views, which we treat as two subsets, i.e., Pediatric (A4C) and Pediatric (PSAX); following its 10-fold protocol, we use folds 1--8 for training, fold 9 for validation, and fold 10 for testing. Each ECHO video is resized to $112\times112$ with 125 frames. Both datasets provide expert-labeled ejection fraction (EF) scores, together with annotated end-diastolic (ED) and end-systolic (ES) frames. 

\begin{table}[h]
\centering
\caption{Details of the CMR and ECHO datasets used in our experiments.}
\setlength{\tabcolsep}{2.0mm}
\resizebox{0.85\linewidth}{!}{
\begin{tabular}{lccccc}
\toprule
\textbf{Datasets} & \makecell{\textbf{Modality}\\\textbf{/ View}} & \textbf{Train.} & \textbf{Val.} & \textbf{Test.} & \makecell{\textbf{Raw Average}\\\textbf{Frame Count}} \\
\midrule
\rowcolor{tablegray}
\multicolumn{6}{c}{\textit{CMR Dataset}} \\
UKBB~\cite{PETERSEN20168} & 3D+t CMR & 3503 & 438 & 438 & 50 \\
\midrule
\rowcolor{tablegray}
\multicolumn{6}{c}{\textit{ECHO Datasets}} \\
EchoNet-Dynamic~\cite{ouyang2020video} & A4C & 7465 & 1288 & 1277 & 177 \\
EchoNet-Pediatric (A4C)~\cite{REDDY2023482} & A4C & 2580 & 336 & 368 & 101 \\
EchoNet-Pediatric (PSAX)~\cite{REDDY2023482} & PSAX & 3559 & 448 & 519 & 103 \\
\bottomrule
\end{tabular}
}
\label{tab:dataset_details}
\end{table}

\begin{table}[t]
\centering
\begin{minipage}[t]{0.47\textwidth}
  \centering
  \caption{Overview of hyperparameter settings for CMR synthesis.}
  \setlength{\tabcolsep}{1.5mm}
  \resizebox{0.90\linewidth}{!}{
  \begin{tabular}{lc}
  \toprule
  \multicolumn{2}{l}{\textbf{Architecture}} \\
  Input dim.   & $50\times16\times16\times16\times4$ \\
  Hidden dim.  & 768  \\
  Num. blocks  & 12    \\
  Num. heads   & 12  \\
  Patch size   & 2   \\
  Params (M)   & 144.30 \\
  FLOPs (G)    & 639.92 \\
  \midrule
  \multicolumn{2}{l}{\textbf{Optimization}} \\
  Batch size        & $4\times4 = 16$  \\
  Optimizer         & AdamW             \\
  Lr                & $5\mathrm{e}{-5}$  \\
  $(\beta_1,\beta_2)$ & $(0.9, 0.95)$   \\
  Lr schedule       & Warmup cosine     \\
  Training steps    & 200k                \\
  EMA decay         & 0.9999        \\
  \midrule
  \multicolumn{2}{l}{\textbf{Interpolants}} \\
  $\alpha_t$  & $1 - t$ \\
  $\sigma_t$  & $t$ \\
  $\omega_t$  & $\sigma_t$ \\
  Training objective   & v-prediction \\
  Sampler              & Euler-Maruyama\\
  Sampling steps       & 250 \\
  \midrule
  \multicolumn{2}{l}{\textbf{CardioFAD}} \\
  Num. keyframes       & 10 \\
  $p(A), p(B)$         & $p(A)= 0.5, p(B)= 0.5$\\
  $\lambda_A, \lambda_B$       & $\lambda_A = 1.5$,  $\lambda_B = 0.5$ \\
  \bottomrule
  \end{tabular}
  }
  \label{syn_settings_CMR}
\end{minipage}\hspace{12pt}
\begin{minipage}[t]{0.47\textwidth}
  \centering
  \caption{Overview of hyperparameter settings for ECHO synthesis.}
  \setlength{\tabcolsep}{1.5mm}
  \resizebox{0.90\linewidth}{!}{
  \begin{tabular}{lc}
  \toprule
  \multicolumn{2}{l}{\textbf{Architecture}} \\
  Input dim.   & $32\times16\times14\times14$ \\
  Hidden dim.  & 768  \\
  Num. blocks  & 12    \\
  Num. heads   & 12  \\
  Patch size   & 2   \\
  Params (M)   & 144.06 \\
  FLOPs (G)    & 157.37 \\
  \midrule
  \multicolumn{2}{l}{\textbf{Optimization}} \\
  Batch size        & $16\times4 = 64$  \\
  Optimizer         & AdamW             \\
  Lr                & $1\mathrm{e}{-4}$  \\
  $(\beta_1,\beta_2)$ & $(0.9, 0.95)$   \\
  Lr schedule       & Warmup cosine     \\
  Training steps    & 200k                \\
  EMA decay         & 0.9999        \\
  \midrule
  \multicolumn{2}{l}{\textbf{Interpolants}} \\
  $\alpha_t$  & $1 - t$ \\
  $\sigma_t$  & $t$ \\
  $\omega_t$  & $\sigma_t$ \\
  Training objective   & v-prediction \\
  Sampler              & Euler-Maruyama\\
  Sampling steps       & 250 \\
  \midrule
  \multicolumn{2}{l}{\textbf{CardioFAD}} \\
  Num. keyframes       & 8 \\
  $p(A), p(B)$         & $p(A)= 0.5, p(B)= 0.5$\\
  $\lambda_A, \lambda_B$       & $\lambda_A = 1.5$,  $\lambda_B = 0.5$ \\
  \bottomrule
  \end{tabular}
  }
  \label{syn_settings_ECHO}
\end{minipage}
\end{table}

\section{More Training Details}

\subsection{Implementation Details}
\label{app:imple}
\textbf{Tokenizer Training}.
Adam is used with batch sizes 4 (CMR) and 32 (ECHO). We set $\sigma$ to 0.2. The weights $\lambda_1$ and $\lambda_2$ are set to 0.1 and 0.05, respectively. The learning rates are \texttt{2e-4}/\texttt{5e-5}. The visual tokenizer, LeanVAE~\cite{Cheng_2025_ICCV}, is optimized with a reconstruction objective $\mathcal{L}_{\mathrm{rec}}$ composed of an $L_{1}$ reconstruction loss, a perceptual loss, an adversarial loss, and a KL regularization term following the original setting. We use LeanVAE~\cite{Cheng_2025_ICCV} with a compression ratio of $4\times 8\times 8$ and 16 latent channels. For CMR, we treat each 3D volume at a cardiac time point as a video by regarding the depth axis as the tokenizer's temporal axis; accordingly, each volume is compressed along the depth, height, and width dimensions with a factor of $4\times 8\times 8$, while the cardiac temporal dimension is kept consistent. For ECHO, we follow the standard video-tokenizer pipeline and compress each video sample along the temporal, height, and width dimensions with a $4\times 8\times 8$ ratio. For temporal alignment, we map both visual and ECG tokens to the same shared temporal length $T$. Specifically, for CMR, the ECG input has length 500 and is tokenized by the ECG ViT with patch size 10, yielding 50 temporal tokens, which matches the temporal length of the visual latent sequence produced by the LeanVAE encoder; similarly, for ECHO, the ECG input has length 128 and is tokenized with patch size 4 into 32 temporal tokens, matching the 32-frame visual latent sequence. Our tokenizer training for ECHO is conducted on the combined EchoNet-Dynamic~\cite{ouyang2020video} and EchoNet-Pediatric~\cite{REDDY2023482} datasets.

\textbf{Diffusion Model Training}.
For the diffusion process, the total batch size is 16/64, and training is performed for 200k steps. We optimize the diffusion model using AdamW~\cite{loshchilov2018decoupled} with a learning rate of \texttt{5e-5}/\texttt{1e-4}. All experiments are conducted on 4 NVIDIA A100 GPUs. We precompute and cache the latents and keyframe scores $q$ on disk before diffusion training for efficiency, which significantly reduces training overhead. We further maintain an exponential moving average (EMA) of model weights with a decay rate of $0.9999$. During sampling, we use the Euler--Maruyama sampler with 250 steps. During inference, keyframes are selected deterministically as the hard top-$K$ indices of the learned importance scores $q$, rather than being stochastically sampled. The detailed hyperparameter setups of our diffusion model are provided in Tables~\ref{syn_settings_CMR} and~\ref{syn_settings_ECHO}.

\textbf{Observation Setup}.
In all sparse-observation settings, the observation is randomly sampled. For \textbf{Volume2Seq}, we observe one 3D volume at a random time index. For \textbf{Slice2Seq}, we observe one 2D slice at a random depth index. For \textbf{Frame2Video}, we observe one frame at a random time index.


\subsection{Details of Two-stage Generation}
\label{app:generation}
\noindent\textbf{Video DiT Architecture.} 
We adopt a factorized Video DiT with alternating spatial and temporal transformer blocks to separately model subject-specific anatomy and cardiac dynamics. The architecture processes both 3D+t CMR and 2D+t ECHO latents using a unified conditioning strategy:

\begin{itemize}
  \item \textit{Input Formulation:} The observed reference latent (volume, slice, or frame) is replicated along the temporal axis and concatenated with the noisy latent along the channel dimension before patch embedding. We apply fixed spatial/temporal positional embeddings alongside modality-specific rotary positional encodings (3D for CMR, 2D for ECHO).
  \item \textit{Spatial Blocks (Anatomy Preservation):} To maintain anatomical consistency, the reference latent is mean-pooled and injected as a global condition via AdaLN. For the Slice2Seq task, the specific slice depth index is concatenated to this spatial condition to provide spatial awareness.
  \item \textit{Temporal Blocks (Motion Generation):} Temporal dynamics are guided by three signals: the observation time index, the keyframe probability $q$, and the physiological time-series embedding (ECG for CMR, pseudo-ECG for ECHO). Their pooled representation is injected globally via AdaLN, while the frame-wise time-series embeddings are additionally injected via cross-attention to provide fine-grained, temporally varying cardiac guidance.
\end{itemize}

\noindent\textbf{Unified Masked Diffusion.} 
We train a single Video DiT using a unified masked diffusion objective for both keyframe generation and non-keyframe interpolation, rather than training separate models. Given a binary keyframe mask $m^{kf}\in\{0,1\}^{T}$, we randomly sample one of two generation modes at each training iteration:

\begin{itemize}
    \item \textit{Mode A (Keyframe Generation):} Supervised using the mask $m=m^{kf}$.
    \item \textit{Mode B (Frame Interpolation):} Supervised using the complementary mask $m=1-m^{kf}$.
\end{itemize}

Following the transport-based diffusion formulation, the network predicts the target velocity field conditioned on all available signals (i.e., the observed reference latent, observation index, time-series embeddings, and keyframe probability). Crucially, the temporal mask $m$ is applied exclusively to the loss function, allowing the same denoiser to supervise different subsets of frames:

\begin{equation}
    \mathcal{L}_{diff}=\lambda_{mode}\text{Mean}[m\odot(v_{\theta}-u_{\tau})^{2}]
\end{equation}

In practice, both modes are sampled with equal probability ($0.5$). To encourage the model to prioritize accurate anchor generation before refining intermediate dynamics, we assign a larger loss weight to Mode A ($\lambda_{A}=1.5$) than to Mode B ($\lambda_{B}=0.5$). This complementary masking strategy enables a single network to efficiently and jointly learn both stages.

\subsection{Pseudo-ECG Construction and Usage for ECHO}
\label{app:pseudo_ecg}
For ECHO datasets lacking native ECG recordings, we construct a surrogate 1D physiological prior (pseudo-ECG) to provide coarse, normalized cardiac-cycle conditioning.

\noindent\textbf{Mathematical Formulation.} We standardize the cardiac cycle to length $N$ (we set $N=128$) with a normalized phase $\phi\in[0,1)$. Setting the ED phase to $\phi_{ED}=0$ and the ES phase to $\phi_{ES}=0.3$ based on standard physiological proportions, we model ventricular volume changes using a smooth cosine interpolation $g(\phi)$:
\begin{equation}
g(\phi)=\begin{cases}\frac{1}{2}[1+\cos(\pi\frac{\phi}{\phi_{ES}})], & 0\le\phi\le\phi_{ES}\\ \frac{1}{2}[1-\cos(\pi\frac{\phi-\phi_{ES}}{1-\phi_{ES}})], & \phi_{ES}<\phi<1\end{cases}
\end{equation}
After min-max normalization, the patient-specific anatomical areas mathematically cancel out, rendering the normalized amplitude exactly equivalent to the canonical template $g(\phi)$. The final 1D pseudo-ECG token at step $n$ is simply concatenated as $e_{n}=[\phi_{n},g(\phi_{n}),\Delta g(\phi_{n})]$.

\noindent\textbf{Patient-Agnostic and Zero-Leakage.} Crucially, because $e_{n}$ reduces to a universal, patient-agnostic template, it requires absolutely no patient-specific annotations (e.g., actual ED/ES frame indices or contour areas) during inference. In our extreme one-shot clinical setting, the model conditions exclusively on this generic periodic signal, strictly ensuring zero temporal information leakage at test time. Moreover, Table~\ref{tab:ablation_condition_ECHO} and Appendix~\ref{app:perturbation} (\textit{Robustness to ECG Perturbations}) provide controlled ablations with degraded/perturbed ECG guidance, showing that the performance gains do not hinge on the exact pseudo-ECG quality and that the proposed AAP-driven alignment remains effective beyond this weak auxiliary cue. A controlled attribution study in Appendix~\ref{app:attribution_temp} further disentangles this temporal guidance, showing that the ECHO gains arise from valid generic phase conditioning and two-stage keyframe anchoring rather than patient-specific ECG information.

\subsection{Details of EF Estimation}
\label{app:EF}
For cardiac functional preservation evaluation, we follow the standard echocardiography setting and treat EF estimation as a downstream regression task that predicts a continuous ejection fraction (EF) value from an input video~\cite{reynaud2025echoflow, reynaud2024echonet}. Specifically, the EF dataset consists of video--label pairs $(V, y)$, where $V$ denotes an echocardiographic sequence and $y$ is the corresponding ground-truth EF score. A regression model $\mathrm{Reg}(\cdot)$ takes the full sequence as input and outputs an estimate $\hat{y}=\mathrm{Reg}(V)$, and is trained by minimizing the mean-squared error, $\mathcal{L}_{\mathrm{reg}}=(y-\hat{y})^2$. We apply the same protocol to evaluate generated videos, using downstream EF performance as a proxy for whether the synthesized dynamics preserve clinically relevant functional cues. Better EF estimation performance therefore suggests stronger cardiac functional consistency in the generated sequences.

\section{Additional Experimental Results}
\subsection{Frozen Visual Tokenizer Validation}
\label{app:frozen_tok}
One potential concern is that the learned keyframes may be partially coupled with the jointly optimized visual latent space. To investigate whether our selected keyframes remain informative under an independently fixed representation, we conduct an additional validation with a frozen visual tokenizer on the EchoNet-Dynamic dataset~\cite{ouyang2020video} in Table~\ref{tab:frozen_tok}. Concretely, we first train our visual encoder, LeanVAE tokenizer~\cite{Cheng_2025_ICCV}, using only the standard reconstruction objective for the same number of training steps as the full model, without the TSP, AAP, or keyframe selector. We then freeze the visual encoder and decoder, and train the ECG encoder, keyframe selector and alignment module on top of this fixed latent space. In this setting, the selector cannot reshape the visual representation to satisfy the TSP objective, and must identify keyframes that are sufficient under an independently learned latent trajectory.

It can be observed from Table~\ref{tab:frozen_tok} that CardioFAD with a frozen tokenizer still outperforms random, uniform, and GT ED/ES keyframes in both latent trajectory reconstruction and synthesis quality. It also clearly improves over the one-stage baseline, indicating that keyframe anchoring remains effective even when the visual latent space is fixed. The joint-tokenizer model further improves performance, suggesting that joint optimization enhances representation compatibility, while the frozen-tokenizer validation demonstrates that the learned keyframes are not merely artifacts of co-adapted latent representations.

\begin{table}[t]
    \centering
    \setlength{\tabcolsep}{2.0mm}
    \caption{\textbf{Frozen visual tokenizer validation.} 
    The visual tokenizer is trained independently and frozen before learning the ECG encoder, selector, alignment module, and diffusion model.}
    \resizebox{0.95\linewidth}{!}{
    \begin{tabular}{lcccccc}
    \toprule
    \textbf{Methods} & Latent MSE$\downarrow$ & PSNR$\uparrow$ & SSIM$\uparrow$ & LPIPS$\downarrow$ & FID$\downarrow$ & FVD$\downarrow$ \\
    \midrule
    Uniform keyframes    & 0.023 & 25.94 & 0.572 & 0.081 & 40.41 & 188.36 \\
    Random keyframes   & 0.027  & 25.17 & 0.554 & 0.091 & 45.59 & 251.03 \\
    GT ED/ES            & 0.020 & 26.20 & 0.572 & 0.083 & 42.76 & 202.41 \\
    \midrule
    \rowcolor{tableblue!80}
    CardioFAD (One-stage)  & -- &  25.14 & 0.556 & 0.089 & 46.82 & 232.17 \\
    \rowcolor{tablepeach!70}
    CardioFAD (Frozen Tokenizer) & 0.018 & 27.56 & 0.592 & 0.070 & 31.84 & 162.93 \\
    \rowcolor{tablepurple!70}
    \textbf{CardioFAD (Joint Tokenizer)} & \textbf{0.014} & \textbf{27.73}  & \textbf{0.596} & \textbf{0.067} & \textbf{26.92} & \textbf{137.70}  \\
    \bottomrule
    \end{tabular}
    }
    \label{tab:frozen_tok}
\end{table}

\begin{table*}[t]
  \centering
  \setlength{\tabcolsep}{2.0mm}
  \renewcommand\arraystretch{1.0}
  \caption{\textbf{Cardiac functional consistency} evaluated by EF estimation for 2D+t ECHO synthesis on EchoNet-Pediatric and its A4C/PSAX subsets~\cite{REDDY2023482}.}
  \resizebox{0.85\linewidth}{!}{
    \begin{tabular}{lcccccc}
      \toprule
      \multirow{2}{*}{\textbf{Methods}}
      & \multicolumn{3}{c}{\textbf{EchoNet-Pediatric (A4C)}}
      & \multicolumn{3}{c}{\textbf{EchoNet-Pediatric (PSAX)}} \\
      \cmidrule(lr){2-4} \cmidrule(lr){5-7}
      & $R^2$$\uparrow$ & MAE$\downarrow$ & RMSE$\downarrow$
      & $R^2$$\uparrow$ & MAE$\downarrow$ & RMSE$\downarrow$ \\
      \midrule
      \rowcolor{tablegray}
      \multicolumn{7}{c}{\textit{General Video Generation}} \\
      VideoComposer~\cite{NEURIPS2023_180f6184} & 0.68 & 4.93 & 6.73 & 0.62 & 4.88 & 7.21  \\
      Tora~\cite{Zhang_2025_CVPR} & 0.66 & 4.89 & 6.86 & 0.69 & 4.59 & 6.52   \\
      \midrule
      \rowcolor{tablegray}
      \multicolumn{7}{c}{\textit{ECHO Video Generation}} \\
      HeartBeat~\cite{zhou2024heartbeat} & 0.71 & 4.62 & 6.29 & 0.65 & 4.40 & 6.94   \\
      EchoNet-Synthetic~\cite{reynaud2024echonet} & 0.70 & 4.57 & 6.44 & 0.68 & 4.19 & 6.60   \\
      ECHOPulse~\cite{li2025echopulse} & 0.73 & 4.39 & 6.07 & 0.72 & 4.06 & 6.15  \\
      \midrule
      \rowcolor{tableblue!80}
      CardioFAD (One-stage) &  0.73 &  4.35 &  6.02 &  0.73 &  4.10 &  6.02 \\
      \rowcolor{tablepurple!70}
      \textbf{CardioFAD (Ours)} &  \textbf{0.75} &  \textbf{4.23} &  \textbf{5.86} &  \textbf{0.76} &  \textbf{3.98} &  \textbf{5.79}  \\
      \bottomrule
    \end{tabular}
  }
  \label{tab:ECHO_clinical_addtion}
\end{table*}

\subsection{Additional Cardiac Functional Evaluation}

\textbf{Functional consistency across pediatric ECHO views.}
We further report cardiac functional consistency evaluation for 2D+t ECHO synthesis on EchoNet-Pediatric and its A4C/PSAX subsets in Table~\ref{tab:ECHO_clinical_addtion}. Across both views, our full model achieves the strongest overall downstream EF estimation performance, while the one-stage variant remains competitive, confirming the benefit of the proposed Focus--Align--Diffuse design. The improvements over both general video generation and ECHO-specific baselines indicate that CardioFAD better preserves EF-related functional information in the synthesized sequences.

\begin{table*}[t]
\centering
\caption{\textbf{EF-stratified functional preservation} on EchoNet-Dynamic.}
\setlength{\tabcolsep}{2.5mm}
\renewcommand{\arraystretch}{1.00}
\resizebox{0.80\linewidth}{!}{
\begin{tabular}{lcccc}
\toprule
\multirow{2}{*}{\textbf{Metric}}
& Reduced
& Mildly reduced
& Preserved-range
& \multirow{2}{*}{All} \\
& $\mathrm{EF}\leq40$
& $40<\mathrm{EF}<50$
& $\mathrm{EF}\geq50$
& \\
\midrule
$n$
& 161 & 125 & 991 & 1,277 \\
\midrule

Real-video MAE $\downarrow$
& 4.81 & 4.49 & 3.76 & 3.96 \\

\rowcolor{tableblue!80}
One-stage MAE $\downarrow$
& 5.84 & 5.29 & 4.80 & 4.98 \\

\rowcolor{tablepurple!70}
\textbf{CardioFAD MAE} $\downarrow$
& \textbf{5.10} & \textbf{4.65} & \textbf{3.96} & \textbf{4.17} \\

Bias vs. GT
& 0.72 & 0.58 & 0.15 & 0.26 \\

Paired Gen--Real MAE $\downarrow$
& 2.13 & 1.79 & 1.40 & 1.53 \\
\bottomrule
\end{tabular}}
\label{tab:ef_stratified}
\end{table*}

\begin{table*}[t]
\centering
\caption{\textbf{Direct motion preservation} on EchoNet-Dynamic.}
\setlength{\tabcolsep}{2.5mm}
\renewcommand{\arraystretch}{1.00}
\resizebox{0.80\linewidth}{!}{
\begin{tabular}{lccc}
\toprule
\multirow{2}{*}{\textbf{Metric}}
& All stratified
& \multirow{2}{*}{Reduced EF}
& High motion \\
& cases
&
& heterogeneity \\
\midrule
$n$
& 60 & 20 & 15 \\
\midrule

\rowcolor{tableblue!80}
Area-curve MAE (One-stage) $\downarrow$
& 0.063 & 0.078 & 0.095 \\

\rowcolor{tablepurple!70}
\textbf{Area-curve MAE (CardioFAD)} $\downarrow$
& \textbf{0.048} & \textbf{0.051} & \textbf{0.067} \\
\midrule

\rowcolor{tableblue!80}
Segmental-motion MAE (One-stage) $\downarrow$
& 0.119 & 0.140 & 0.182 \\

\rowcolor{tablepurple!70}
\textbf{Segmental-motion MAE (CardioFAD)} $\downarrow$
& \textbf{0.097} & \textbf{0.116} & \textbf{0.141} \\
\bottomrule
\end{tabular}}
\label{tab:direct_motion}
\end{table*}

\textbf{Pathology-specific motion patterns.}
We further examine whether CardioFAD preserves EF-related functional differences and regional motion patterns on EchoNet-Dynamic.
We first divide all 1,277 test subjects into reduced ($\mathrm{EF}\leq40$), mildly reduced ($40<\mathrm{EF}<50$), and preserved-range ($\mathrm{EF}\geq50$) groups.
Using the same frozen EF regressor, Table~\ref{tab:ef_stratified} reports the real-video MAE, generated-video MAE, signed bias, and paired generated--real MAE. CardioFAD improves over the one-stage model in every EF group, while maintaining only modest bias, supporting the preservation of EF-related severity differences.

We further evaluate direct motion consistency on a stratified subset of 60 subjects (20 per EF group) using a frozen LV segmenter.
The 15 cases with the highest real-video segmental-motion heterogeneity are additionally evaluated as a challenging subset.
Area-curve MAE measures the discrepancy between ED-normalized LV area trajectories over the cardiac cycle, while segmental-motion MAE compares six A4C endocardial-segment shortening profiles from ED to ES. CardioFAD reduces both errors across all evaluated subsets in Table~\ref{tab:direct_motion}, including reduced-EF and high-heterogeneity cases. The consistent gains suggest improved preservation of global contraction--relaxation and heterogeneous regional motion.

\begin{table}[t]
    \centering
    \setlength{\tabcolsep}{2.0mm}
    \caption{\textbf{Multi-seed robustness analysis on EchoNet-Dynamic.} Results are reported as mean ± standard deviation over three independent random seeds.}
    \resizebox{0.85\linewidth}{!}{
    \begin{tabular}{lccccc}
    \toprule
    \textbf{Methods} & PSNR$\uparrow$ & SSIM$\uparrow$ & LPIPS$\downarrow$ & FID$\downarrow$ & FVD$\downarrow$ \\
    \midrule
    ECHOPulse~\cite{li2025echopulse}    & 25.60\stdgray{0.19} & 0.567\stdgray{0.003} & 0.081\stdgray{0.004} & 37.24\stdgray{2.5} & 194.19\stdgray{11.5} \\
    \midrule
    \rowcolor{tableblue!80}
    CardioFAD (One-stage)   &  25.17\stdblue{0.16} & 0.558\stdblue{0.003} & 0.090\stdblue{0.003} & 46.52\stdblue{2.9} & 235.08\stdblue{9.8} \\
    \rowcolor{tablepurple!70}
    \textbf{CardioFAD (Ours)}  & \textbf{27.68}\stdpurple{0.20}  & \textbf{0.595}\stdpurple{0.004} & \textbf{0.068}\stdpurple{0.004} & \textbf{26.48}\stdpurple{2.2} & \textbf{139.54}\stdpurple{12.8}  \\
    \bottomrule
    \end{tabular}
    }
    \label{tab:multi-seed}
\end{table}

\subsection{Multi-seed Robustness Analysis}
\label{app:multi-seed}

To assess robustness to training randomness, we train and evaluate all methods using three independent random seeds on EchoNet-Dynamic. As reported in Table~\ref{tab:multi-seed}, CardioFAD consistently outperforms ECHOPulse~\cite{li2025echopulse} and the one-stage variant across all metrics. In particular, its 2.08 dB PSNR gain over ECHOPulse is substantially larger than the across-seed variation, with a two-sided Welch’s t-test yielding \(p<0.001\). These results confirm the robustness of our improvements and the benefit of the two-stage design.

\subsection{Additional Ablation Studies}
\label{app:ablation}

\begin{table}[t]
  \centering
  \setlength{\tabcolsep}{2.0mm}
  \renewcommand\arraystretch{1.0}
  \caption{\textbf{Principle-guided keyframe learning ablation} for CMR Volume2Seq synthesis.}
  \resizebox{0.70\linewidth}{!}{
    \begin{tabular}{lccccc}
      \toprule
      \textbf{Methods} & PSNR$\uparrow$ & SSIM$\uparrow$ & LPIPS$\downarrow$ & FID$\downarrow$ & FVD$\downarrow$ \\
      \midrule
      \rowcolor{tablegray}
      \multicolumn{6}{c}{\textit{Keyframe Learning Principle}} \\
      w/o TSP              & 29.07 & 0.836 & 0.098 & 104.53 & 1226.37   \\
      w/o AAP              & 30.53 & 0.861 & 0.082 & 81.63 & 910.04  \\
      w/o TSP \& w/o AAP              & 28.21 & 0.827 & 0.111 & 134.54 & 1507.13   \\
      \midrule
      \rowcolor{tablegray}
      \multicolumn{6}{c}{\textit{Keyframe Selection Strategy}} \\
      w/ uniform keyframes       & 29.86 & 0.852 & 0.094 & 97.41 & 1028.59 \\
      w/ random keyframes        & 28.63 & 0.834 & 0.105 & 121.86 & 1379.10 \\
      \rowcolor{tablepurple!70}
      \textbf{CardioFAD}       &  \textbf{32.41}  &  \textbf{0.885} &  \textbf{0.064} &  \textbf{59.96} &  \textbf{736.01}  \\
      \bottomrule
    \end{tabular}
    } 
  \label{tab:ablation_principles_CMR}
\end{table}

\begin{table*}[t]
\centering
\begin{minipage}[t]{0.48\textwidth}
  \centering
  \setlength{\tabcolsep}{1.5mm}
  \renewcommand\arraystretch{1.0}
  \caption{\textbf{Ablating the number of keyframes} for CMR Volume2Seq synthesis.}
  \resizebox{1.0\linewidth}{!}{
    \begin{tabular}{lccccc}
      \toprule
      \textbf{Methods} & PSNR$\uparrow$ & SSIM$\uparrow$ & LPIPS$\downarrow$ & FID$\downarrow$ & FVD$\downarrow$ \\
      \midrule
      $K=2$  & 30.29 & 0.863 & 0.083 & 86.26 & 1063.45 \\
      $K=5$  & 30.90 & 0.867 & 0.079 & 80.73 & 945.92 \\
      $K=8$  & 32.27 & 0.882 & 0.068 & 65.93 & 825.14 \\
      \rowcolor{tablepurple!70}
      $\mathbf{K=10}$   &  \textbf{32.41}  &  \textbf{0.885} &  \textbf{0.064} &  \textbf{59.96} &  \textbf{736.01}  \\
      $K=15$ & 31.16 & 0.873 & 0.073 & 84.67 & 907.22 \\
      \bottomrule
    \end{tabular}
  }
  \label{tab:number_keyframe_CMR}
\end{minipage}\hfill
\begin{minipage}[t]{0.48\textwidth}
  \centering
  \setlength{\tabcolsep}{1.0mm}
  \renewcommand\arraystretch{1.05}
  \caption{\textbf{Ablating the number of keyframes} on EchoNet-Dynamic.}
  \resizebox{\linewidth}{!}{
    \begin{tabular}{lccccc}
      \toprule
      \textbf{Methods} & PSNR$\uparrow$ & SSIM$\uparrow$ & LPIPS$\downarrow$ & FID$\downarrow$ & FVD$\downarrow$ \\
      \midrule
      $K=2$  & 26.18 & 0.574 & 0.077 & 36.30 & 173.79 \\
      $K=4$  & 27.11 & 0.583 & 0.072 & 30.64 & 148.27 \\
      \rowcolor{tablepurple!70}
      $\mathbf{K=8}$   &  \textbf{27.73}  &  \textbf{0.596} &  \textbf{0.067} &  \textbf{26.92} &  \textbf{137.70}  \\
      $K=12$ & 27.36 & 0.583 & 0.069 & 28.24 & 161.08 \\
      \bottomrule
    \end{tabular}
  }
  \label{tab:number_keyframe_ECHO}
\end{minipage}
\end{table*}

\begin{table*}[t]
\centering
\begin{minipage}[t]{0.49\textwidth}
  \centering
  \setlength{\tabcolsep}{1.0mm}
  \renewcommand\arraystretch{1.0}
  \caption{\textbf{Condition ablation} for CMR Volume2Seq synthesis.}
  \resizebox{\linewidth}{!}{
    \begin{tabular}{lccccc}
      \toprule
      \textbf{Methods} & PSNR$\uparrow$ & SSIM$\uparrow$ & LPIPS$\downarrow$ & FID$\downarrow$ & FVD$\downarrow$ \\
      \midrule
      w/o alignment   & 30.89 & 0.864 & 0.081 & 77.85 & 913.72 \\
      w/o ECG condition  & 29.24 & 0.834 & 0.099 & 107.39 & 1281.46 \\
      w/o keyframe prob.    & 31.12 & 0.869 & 0.074 & 80.92 & 866.04 \\
      w/o reference index   & 31.37 & 0.873 & 0.072 & 77.05 & 901.53 \\
      \rowcolor{tablepurple!70}
      \textbf{CardioFAD}    &  \textbf{32.41}  &  \textbf{0.885} &  \textbf{0.064} &  \textbf{59.96} &  \textbf{736.01}  \\
      \bottomrule
    \end{tabular}
  }
  \label{tab:ablation_condition_CMR}
\end{minipage}\hfill
\begin{minipage}[t]{0.49\textwidth}
  \centering
  \setlength{\tabcolsep}{1.0mm}
  \renewcommand\arraystretch{1.0}
  \caption{\textbf{Condition ablation} on EchoNet-Dynamic.}
  \resizebox{\linewidth}{!}{
    \begin{tabular}{lccccc}
      \toprule
      \textbf{Methods} & PSNR$\uparrow$ & SSIM$\uparrow$ & LPIPS$\downarrow$ & FID$\downarrow$ & FVD$\downarrow$ \\
      \midrule
      w/o alignment            & 25.71 & 0.562 & 0.080 & 39.54 & 209.67 \\
      w/o ECG condition        & 25.26 & 0.559 & 0.088 & 45.98 & 243.62 \\
      w/o keyframe prob.       & 26.82 & 0.584 & 0.073 & 32.28 & 169.74 \\
      w/o reference index      & 26.27 & 0.576 & 0.079 & 40.46 & 188.75 \\
      \rowcolor{tablepurple!70}
      \textbf{CardioFAD}       &  \textbf{27.73}  &  \textbf{0.596} &  \textbf{0.067} &  \textbf{26.92} &  \textbf{137.70}  \\
      \bottomrule
    \end{tabular}
  }
  \label{tab:ablation_condition_ECHO}
\end{minipage}
\end{table*}



\textbf{Principle-guided Keyframe Learning for CMR Synthesis}. 
We conduct an additional ablation on principle-guided keyframe learning for CMR Volume2Seq synthesis on the UKBB dataset (Table~\ref{tab:ablation_principles_CMR}). Removing either the TSP or the AAP consistently degrades performance across all metrics, and disabling both leads to the most pronounced deterioration, indicating that the two principles are complementary. We further compare different keyframe selection strategies and find that uniform or random keyframes perform notably worse than our learned keyframes. Overall, these results verify that principled keyframe learning is crucial for achieving high-fidelity and temporally consistent synthesis under extreme sparsity.

\textbf{Effect of Number of Keyframes $K$}. 
We further analyze how the number of learned keyframes affects CardioFAD in Tables~\ref{tab:number_keyframe_CMR} and~\ref{tab:number_keyframe_ECHO}. As shown, a moderate number of keyframes provides the best trade-off between trajectory coverage and focused anchor selection. On CMR Volume2Seq synthesis, performance steadily improves as $K$ increases from 2 to 10, with $K{=}10$ achieving the best results across all metrics. However, further increasing $K$ to 15 degrades both fidelity and temporal consistency, indicating that too many anchors may introduce redundancy and dilute the focus on motion-salient phases. A similar trend is observed on EchoNet-Dynamic~\cite{ouyang2020video}, where $K{=}8$ yields the best overall performance, while both smaller settings ($K{=}2,4$) and a larger setting ($K{=}12$) lead to inferior results. These results support our design choice of using compact keyframe sets, with $K{=}10$ for CMR and $K{=}8$ for ECHO, to capture sufficient cardiac dynamics without weakening the anchor-based generation process.

\textbf{Effect of Conditioning Guidance}. 
We further ablate the conditioning signals used in our video DiT in Tables~\ref{tab:ablation_condition_CMR} and~\ref{tab:ablation_condition_ECHO}. It can be observed that each condition contributes to the generation quality, and removing any component leads to degradation across both CMR and ECHO settings. Specifically, removing ECG conditioning causes the most severe drop, especially in FID and FVD, indicating that physiological time-series guidance is essential for resolving temporal ambiguity under sparse visual observations. Removing cross-modal alignment also degrades performance, verifying that ECG signals cannot be fully exploited without explicitly aligning them with the visual cardiac trajectory. In addition, removing keyframe probabilities weakens the model's awareness of motion-salient anchors, while removing the reference-frame index makes it harder to localize the sparse observation within the cardiac cycle. To sum up, these ablation results confirm that CardioFAD benefits from jointly conditioning the diffusion model on physiological signals, aligned cross-modal representations, keyframe saliency, and observation position, rather than relying on any single source of guidance.

\subsection{Qualitative Results of Ablation Studies}
\begin{figure*}[t]
\centering
\includegraphics[width=1.00\linewidth]{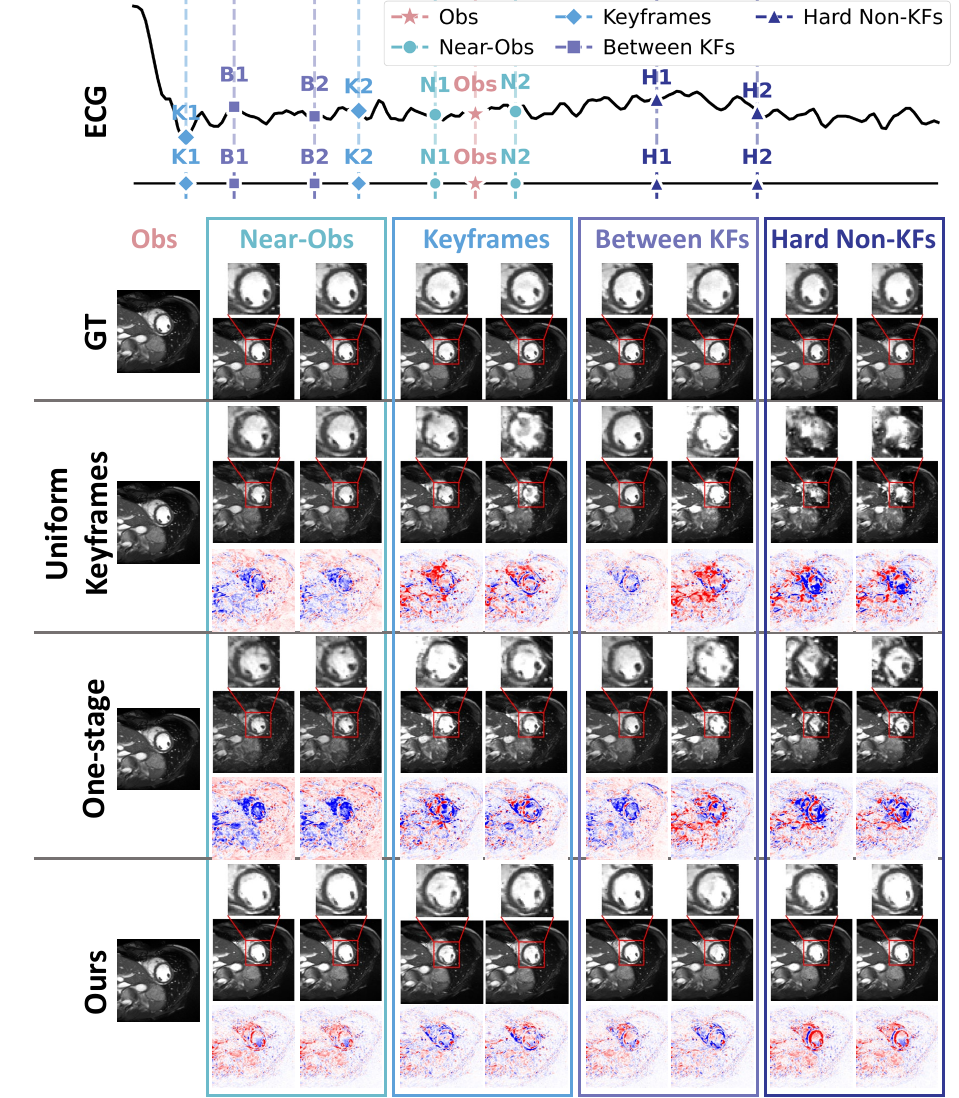}
\caption{\textbf{Qualitative comparisons with variants on CMR.}}
\label{qualitative_ablation_CMR}
\end{figure*}

\begin{figure*}[t]
\centering
\includegraphics[width=1.00\linewidth]{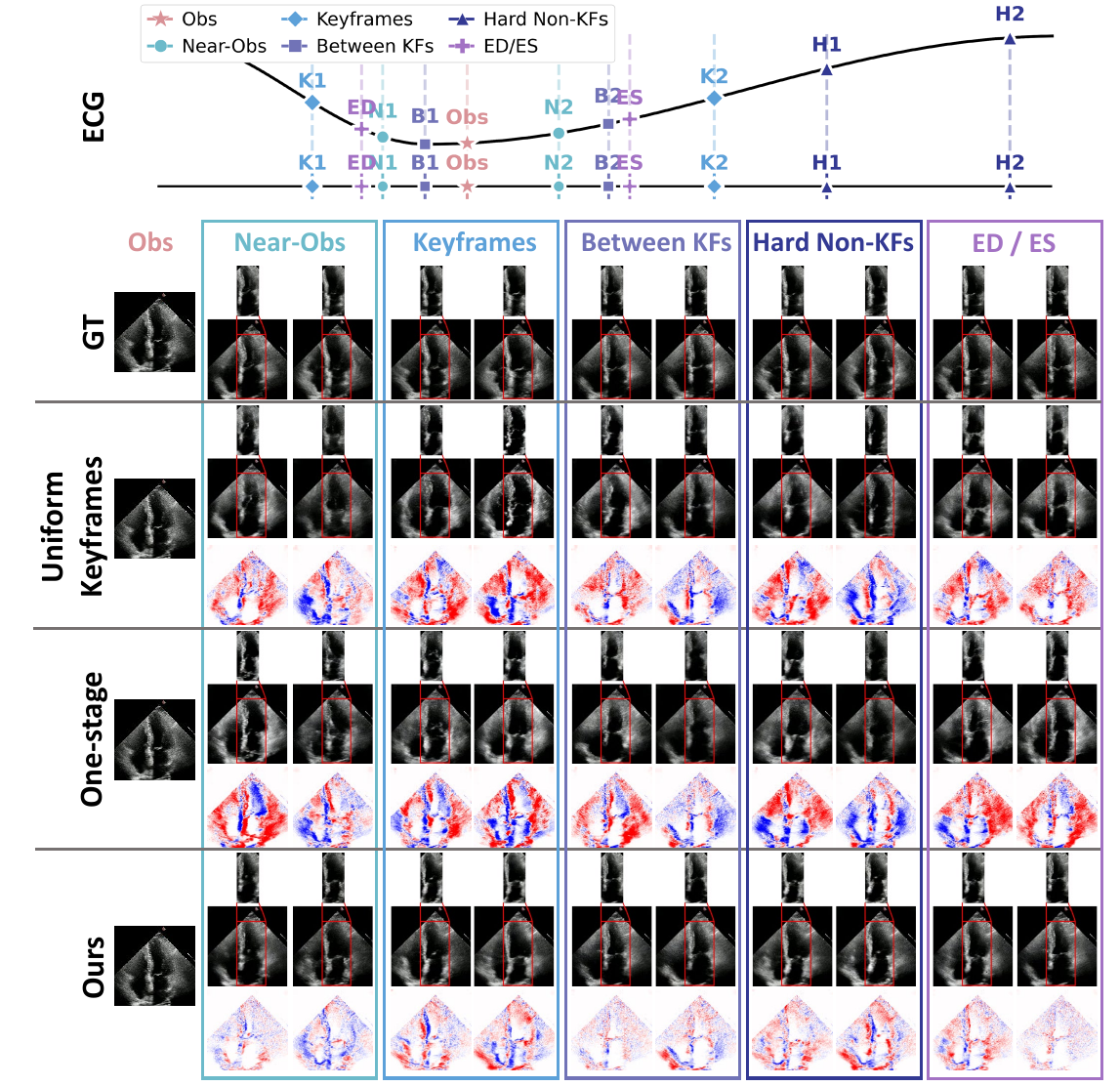}
\caption{\textbf{Qualitative comparisons with variants on ECHO.}}
\label{qualitative_ablation_ECHO}
\end{figure*}

Figs.~\ref{qualitative_ablation_CMR} and~\ref{qualitative_ablation_ECHO} compare our CardioFAD with the one-stage variant and uniform keyframe sampling under the same sparse observation. Across both CMR and ECHO, CardioFAD yields sharper anatomy and more coherent temporal evolution, especially for frames between keyframes and the hard non-keyframes where motion is most ambiguous. In contrast, uniform keyframes and the one-stage variant exhibit stronger blurring and localized artifacts, which is reflected by larger residuals in the difference maps.



\begin{table}[t]
\centering
\setlength{\tabcolsep}{2.0mm}
\caption{\textbf{Attribution of temporal guidance on UKBB and EchoNet-Dynamic.} We vary only the temporal condition under otherwise matched settings to disentangle patient-specific ECG information, generic phase guidance, and two-stage keyframe anchoring. }
\resizebox{\linewidth}{!}{
\begin{tabular}{llcccccc}
\toprule
\textbf{Dataset} & Temporal condition & \makecell{Patient-\\specific} & PSNR$\uparrow$ & SSIM$\uparrow$ & LPIPS$\downarrow$ & FID$\downarrow$ & FVD$\downarrow$ \\
\midrule

\multirow{4}{*}{\textbf{UKBB}}
& \cellcolor{tablepurple!70}\textbf{Paired real ECG}             & \cellcolor{tablepurple!70}$\checkmark$ & \cellcolor{tablepurple!70}\textbf{32.41} & \cellcolor{tablepurple!70}\textbf{0.885} & \cellcolor{tablepurple!70}\textbf{0.064} & \cellcolor{tablepurple!70}\textbf{59.96} & \cellcolor{tablepurple!70}\textbf{736.01} \\
& Subject-shuffled real ECG   & $\times$     & 31.12 & 0.870 & 0.076 & 91.48 & 958.23 \\
& Canonical training ECG      & $\times$     & 30.74 & 0.862 & 0.083 & 98.77 & 1046.28 \\
& Null ECG                    & $\times$     & 29.24 & 0.834 & 0.099 & 107.39 & 1281.46 \\

\midrule

\multirow{4}{*}{\textbf{EchoNet-Dynamic}}
& \cellcolor{tablepurple!70}\textbf{Correct phase template}      & \cellcolor{tablepurple!70}$\times$     & \cellcolor{tablepurple!70}\textbf{27.73} & \cellcolor{tablepurple!70}\textbf{0.596} & \cellcolor{tablepurple!70}\textbf{0.067} & \cellcolor{tablepurple!70}\textbf{26.92} & \cellcolor{tablepurple!70}\textbf{137.70} \\
& Random-offset template      & $\times$     & 25.59 & 0.563 & 0.085 & 37.93 & 217.08 \\
& Null template               & $\times$     & 25.26 & 0.559 & 0.088 & 45.98 & 243.62 \\
& Correct template, one-stage & $\times$     & 25.14 & 0.556 & 0.089 & 46.82 & 232.17 \\

\bottomrule
\end{tabular}
}
\label{tab:temporal_guidance}
\end{table}

\subsection{Attribution of Temporal Guidance}
\label{app:attribution_temp}

As reported in Table~\ref{tab:temporal_guidance}, we conduct a controlled attribution study by varying only the temporal condition under otherwise matched settings. On UKBB~\cite{PETERSEN20168}, paired real ECG consistently outperforms subject-shuffled, canonical, and null ECG, indicating that generic temporal structure is beneficial while correct patient-specific correspondence provides additional gains. On EchoNet-Dynamic~\cite{ouyang2020video}, the correct phase template outperforms both random-offset and null templates, while the two-stage model further improves over its matched one-stage counterpart. These results indicate that patient-specific ECG provides additional value on UKBB, whereas the ECHO gains arise from valid generic phase conditioning together with two-stage keyframe anchoring.

\begin{table}[t]
\centering
\setlength{\tabcolsep}{1.8mm}
\caption{\textbf{Conditional fidelity--diversity analysis on EchoNet-Dynamic.} Eight samples are generated for each fixed sparse observation by varying only the rectified-flow sampling seed.}
\resizebox{\linewidth}{!}{
\begin{tabular}{lccccccc}
\toprule
\textbf{Methods}
& Avg. LPIPS$\downarrow$
& \makecell{Best-of-8\\LPIPS$\downarrow$}
& \makecell{Pairwise\\LPIPS}
& \makecell{Norm. motion\\diversity}
& \makecell{Uncertainty--\\error$\uparrow$}
& EF MAE$\downarrow$
& \makecell{Within-case\\EF std.$\downarrow$} \\
\midrule
\rowcolor{tableblue!80}
CardioFAD (One-stage)
& 0.091
& 0.085
& 0.035
& 0.40
& 0.29
& 5.03
& 2.21 \\

\rowcolor{tablepurple!70}
\textbf{CardioFAD (Ours)}
& \textbf{0.070}
& \textbf{0.064}
& \textbf{0.025}
& \textbf{0.27}
& \textbf{0.50}
& \textbf{4.21}
& \textbf{1.35} \\
\bottomrule
\end{tabular}
}
\label{tab:fidelity_diversity}
\end{table}

\subsection{Conditional Fidelity--Diversity Analysis}
\label{app:fidelity-diversity}

One-shot cardiac synthesis is inherently one-to-many, and our notion of \emph{recoverability} refers to trajectory reconstruction sufficiency rather than unique identification of the underlying motion. 

\textbf{Experimental Setup.}
To examine the resulting conditional uncertainty, we evaluate 100 EchoNet-Dynamic~\cite{ouyang2020video} test cases and generate eight samples for each fixed sparse observation. We keep the observed frame, observation index, pseudo-ECG, and hard top-$K$ keyframes fixed, while varying only the rectified-flow sampling seed.

\textbf{Evaluation Metrics.}
We report Avg. LPIPS for average sample fidelity, Best-of-8 LPIPS for coverage of the held-out realization, and Pairwise LPIPS and normalized motion diversity for appearance and motion variation across samples. Uncertainty--error $\rho$ denotes the Spearman correlation between frame-wise sample disagreement and reconstruction error, while EF MAE and within-case EF standard deviation evaluate functional fidelity and stability.

As reported in Table~\ref{tab:fidelity_diversity}, CardioFAD improves both average and best-of-eight fidelity while reducing unconstrained appearance and motion variation. Its uncertainty--error correlation increases from 0.29 to 0.50, while EF MAE and within-case EF variability decrease to 4.21 and 1.35, respectively, indicating that the remaining stochastic variation is better localized to difficult frames and functionally more stable. Overall, CardioFAD prioritizes faithful and phase-consistent reconstruction over maximal diversity rather than attempting to model the complete conditional posterior.

\subsection{Sensitivity to the Observed-Frame Phase}

To examine sensitivity to the location of the sparse observation within the cardiac cycle, we conduct a phase-stratified evaluation on EchoNet-Dynamic~\cite{ouyang2020video}. All videos are temporally resampled to 125 frames, and we use the same trained checkpoint and inference protocol while changing only the observed frame and its temporal index. ED/ES annotations are used only to define the phase groups for this analysis and are never provided to the model. Using circular temporal indexing, we define four observation settings:

\begin{itemize}
    \item \textbf{Near ED}: frames within $\pm2$ frames of the annotated ED frame, forming a five-frame window.
    \item \textbf{Near ES}: frames within $\pm2$ frames of the annotated ES frame, also forming a five-frame window.
    \item \textbf{Contraction}: intermediate frames between the near-ED and near-ES windows along the ED-to-ES trajectory.
    \item \textbf{Relaxation}: intermediate frames between the near-ES window and the next near-ED window along the ES-to-ED trajectory.
\end{itemize}

Table~\ref{tab:observed_phase} indicates that near-ED and near-ES observations perform slightly better than intermediate contraction and relaxation phases, consistent with motion-extremal states providing stronger trajectory constraints. Nevertheless, the maximum variation is limited to only 0.24 dB in PSNR, 15.9 in FVD, and 0.18 in EF MAE. These results indicate that CardioFAD is only mildly sensitive to the observed-frame phase, with phase conditioning and two-stage anchoring largely compensating for less informative observation positions.

\begin{table}[t]
\centering
\caption{\textbf{Sensitivity to the observed-frame phase} on EchoNet-Dynamic under matched inference settings.}
\resizebox{0.60\linewidth}{!}{
\begin{tabular}{lccc}
\toprule
Observed-frame phase & PSNR$\uparrow$ & FVD$\downarrow$ & EF MAE$\downarrow$ \\
\midrule
Near ED              & \textbf{27.92} & \textbf{128.92} & 4.10 \\
Near ES              & 27.84 & 135.04 & \textbf{4.08} \\
Contraction           & 27.72 & 140.57 & 4.23 \\
Relaxation            & 27.68 & 144.82 & 4.26 \\
Random observation    & 27.73 & 137.70 & 4.17 \\
\bottomrule
\end{tabular}
}
\label{tab:observed_phase}
\end{table}



\subsection{Robustness to ECG Perturbations}
\label{app:perturbation}
\noindent\textbf{ECG Perturbations for CMR Synthesis}.
In real-world clinical settings, auxiliary ECG signals can be corrupted by measurement noise, missing leads, or temporal desynchronization. To assess the intrinsic robustness of our method, we evaluate models trained only on clean data under controlled ECG perturbations only at inference time on UKBB.

\noindent\textit{(i) Signal Noise.}
We inject Gaussian noise into the ECG condition with probability $p\in\{0.1,0.3,0.5,0.7,0.9\}$ (with a fixed noise magnitude $\sigma$, or equivalently matched to a target SNR), simulating stochastic acquisition artifacts.
Concretely, for each time index $t$, we apply
$\tilde{\mathbf{e}}_{t}=\mathbf{e}_{t}+m_{t}\cdot\sigma\boldsymbol{\epsilon}_{t}$,
where $m_{t}\sim\mathrm{Bernoulli}(p)$ and $\boldsymbol{\epsilon}_{t}\sim\mathcal{N}(\mathbf{0},\mathbf{I})$.

\noindent\textit{(ii) Missing Leads.}
To mimic sensor detachment, we randomly drop $L\in\{1,2,4,6,8\}$ leads by zero-masking the corresponding channels, while keeping the input dimensionality unchanged.

\noindent\textit{(iii) Phase Jitter.}
To emulate hardware synchronization lag, we temporally shift the ECG sequence by $\delta\in\{1,2,4,6,8\}$ frames (using circular shift or padding-based shift), introducing controlled misalignment between ECG and image dynamics.

\begin{figure*}[t]
\centering
\includegraphics[width=1.0\linewidth]{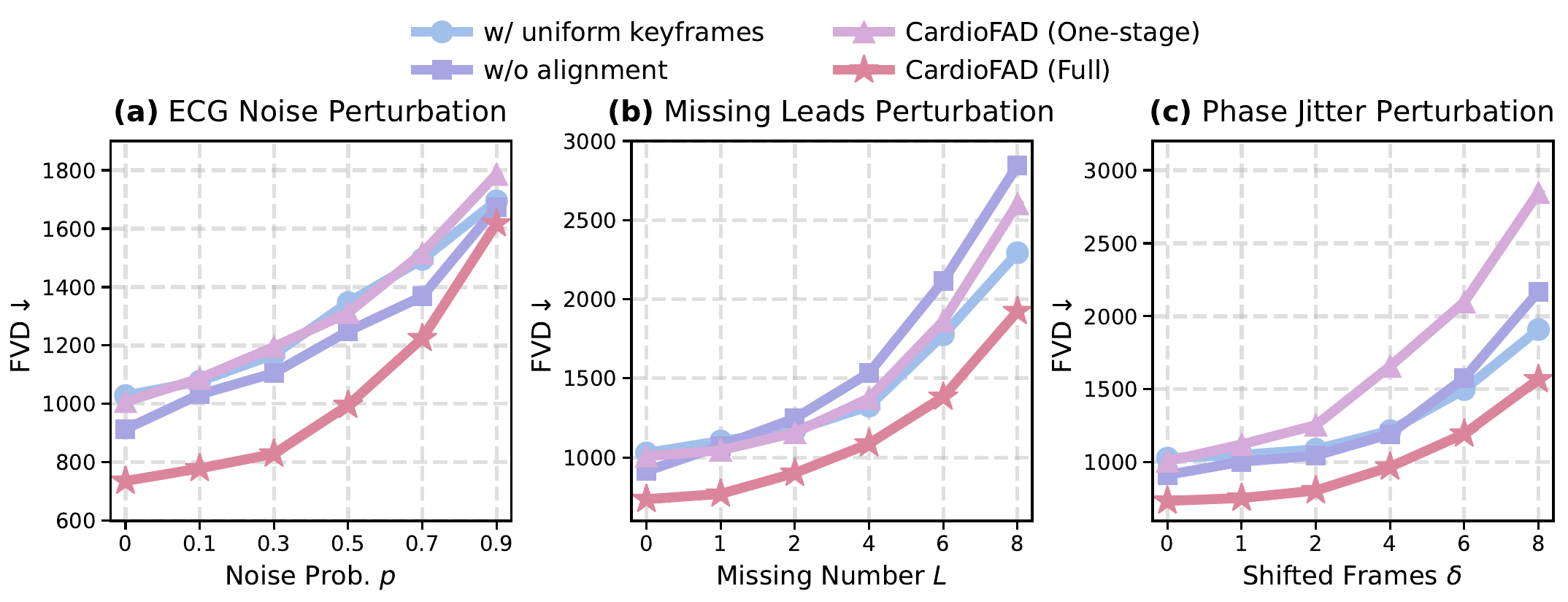}
\caption{\textbf{Robustness to ECG perturbations on UKBB (CMR).} FVD (lower is better) under three inference-time ECG corruptions: Gaussian noise injection with probability $p$, random lead dropout with $L$ missing leads, and temporal phase jitter with $\delta$ shifted frames. CardioFAD (Full) degrades most gracefully and remains consistently best across perturbation severities.}
\label{CMR_perturbation}
\end{figure*}

As shown in Fig.~\ref{CMR_perturbation}, while all methods degrade under ECG perturbations, CardioFAD degrades most gracefully. The performance gap is especially pronounced under missing leads and phase jitter. This resilience indicates that our AAP acts as a robust soft constraint; rather than overfitting to exact channel-wise amplitudes or high-frequency fluctuations, CardioFAD successfully anchors generation using macro-level rhythm and phase cues, supporting robustness to the evaluated conditioning corruptions.

\begin{figure*}[t]
\centering
\includegraphics[width=1.0\linewidth]{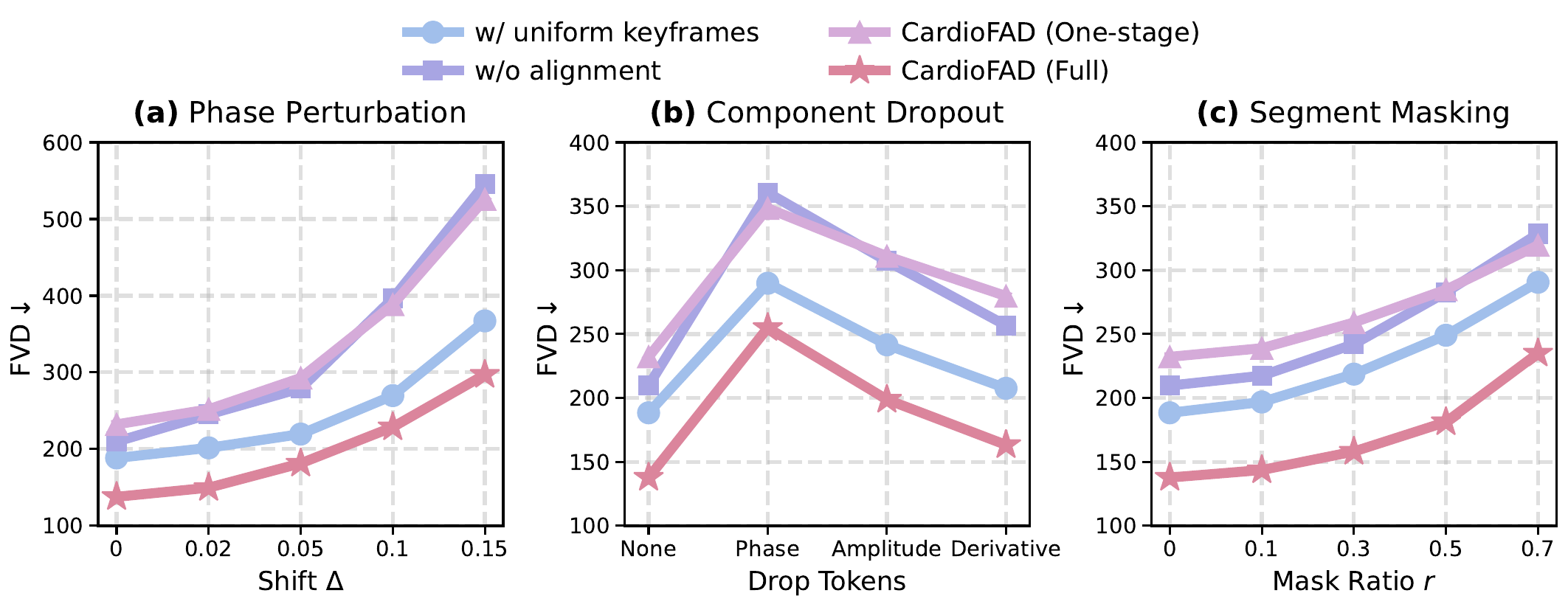}
\caption{\textbf{Robustness to pseudo-ECG perturbations on EchoNet-Dynamic.} FVD (lower is better) under inference-time perturbations of the pseudo-ECG: global phase shifts $\Delta$, component dropout (zero-masking phase/amplitude/derivative), and random segment masking with ratio $r$. CardioFAD (Full) remains consistently best and degrades most gracefully across perturbation severities.}
\label{ECHO_perturbation}
\end{figure*}

\noindent\textbf{Pseudo-ECG Perturbations for ECHO Synthesis}.
For ECHO, ECG is unavailable and we condition the model on a lightweight pseudo-ECG token sequence. To evaluate robustness to imperfect surrogate signals, we perturb the pseudo-ECG only at inference time on EchoNet-Dynamic, while keeping all models trained on clean pseudo-ECG unchanged.

\noindent\textit{(i) Phase Perturbation.}
We apply a global phase shift $\Delta\in\{0.02,0.05,0.10,0.15\}$ to the phase token, simulating synchronization errors and phase ambiguity in template-based conditioning.

\noindent\textit{(ii) Component Dropout.}
We ablate the pseudo-ECG components by dropping one channel at a time (phase / amplitude / derivative) via zero-masking, testing which cues the model relies on most.

\noindent\textit{(iii) Segment Masking.}
We randomly mask a portion of pseudo-ECG tokens with mask ratio $r\in\{0.1,0.3,0.5,0.7\}$ to mimic missing or incomplete conditioning sequences.

Similarly, for pseudo-ECG perturbations (Fig.~\ref{ECHO_perturbation}), CardioFAD consistently exhibits the slowest degradation. While phase alterations severely impact baselines, our model remains substantially more stable. Furthermore, the relatively mild degradation caused by dropping amplitude/derivative channels or masking segments confirms that CardioFAD primarily leverages coarse phase cues rather than fine-grained waveform details, highlighting its robustness under imperfect surrogate conditions.

\subsection{Efficiency Analysis}
\label{app:efficiency}
To validate the practical viability of our proposed CardioFAD, we analyze its computational footprint across both inference and tokenizer training, demonstrating that our principle-guided framework achieves substantial generative quality improvements with negligible memory overhead and acceptable runtime costs.

\noindent\textbf{Inference Cost}. 
In terms of inference efficiency (Table~\ref{inference_cost}), CardioFAD introduces a moderate runtime overhead compared with the one-stage baseline, while maintaining nearly identical memory usage. On both CMR and ECHO, the end-to-end latency is approximately doubled, which is consistent with our two-stage sampling design that sequentially performs keyframe generation and sequence completion. Given the clear performance gains of CardioFAD over the one-stage baseline, this increased runtime with virtually no additional VRAM footprint reflects a reasonable quality--efficiency trade-off. Notably, to reduce preprocessing overhead and better reflect the actual generation cost, all latent representations are pre-encoded and cached locally before timing.

\noindent\textbf{Wall-clock Cost of Tokenizer Training}.
As shown in Table~\ref{Wall-clock_Cost}, across both CMR and ECHO, incorporating TSP introduces only a minor computational overhead, as the additional parameters/FLOPs mainly come from a lightweight UNet used to reconstruct the full cardiac latent, leading to only marginal increases in per-epoch time and total GPU hours. In contrast, AAP does not add any learnable components and thus keeps the model size and FLOPs essentially unchanged. Despite these minimal training-cost changes, the full model achieves consistently better synthesis quality, yielding higher PSNR and substantially lower FVD compared with the corresponding ablations. Overall, these results indicate a favorable quality--efficiency trade-off, where notable quality gains are obtained with negligible additional wall-clock training burden.

\begin{table}[t]
\centering
\caption{\textbf{Inference cost comparison between the one-stage baseline and CardioFAD on CMR and ECHO.}
We report end-to-end latency per sample and peak VRAM under standardized settings (batch size $=1$, single A100 GPU, same sampling steps, and mixed precision).}
\label{tab:inference_cost}
\setlength{\tabcolsep}{2.0mm}
\resizebox{0.90\linewidth}{!}{
\begin{tabular}{lcccc}
\toprule
\multirow{2}{*}{\textbf{Methods}} 
& \multicolumn{2}{c}{CMR (UKBB)} 
& \multicolumn{2}{c}{ECHO (EchoNet-Dynamic)} \\
\cmidrule(lr){2-3} \cmidrule(lr){4-5}
& Latency (s)$\downarrow$ & Peak VRAM (GB)$\downarrow$
& Latency (s)$\downarrow$ & Peak VRAM (GB)$\downarrow$ \\
\midrule
\rowcolor{tableblue!80}
CardioFAD (One-stage) &  6.19 &  14.82 &  3.34 &  4.91 \\
\rowcolor{tablepurple!70}
CardioFAD (Ours) &  12.27 &  14.84 &  6.62 &  4.93 \\
\bottomrule
\end{tabular}
}
\label{inference_cost}
\end{table}

\begin{table*}[t]
\centering
\caption{\textbf{Tokenizer training efficiency vs. downstream synthesis quality.} We compare principle ablations in tokenizer training against the full model. Synthesis quality is reported as the average across the two CMR evaluation settings and across the three ECHO datasets, respectively.}
\label{tab:tokenizer_efficiency_quality}
\setlength{\tabcolsep}{2.0mm}
\resizebox{0.90\textwidth}{!}{
\begin{tabular}{llcccc|cc}
\toprule
\multirow{2}{*}{\textbf{Data}} & \multirow{2}{*}{\textbf{Methods}}
& \multicolumn{4}{c|}{Tokenizer Training Cost}
& \multicolumn{2}{c}{Synthesis Quality} \\
\cmidrule(lr){3-6} \cmidrule(lr){7-8}
&  & {\makecell{Params\\(M) $\downarrow$}}  & {\makecell{FLOPs\\(G) $\downarrow$}} & {\makecell{{Time/Epoch}\\(s) $\downarrow$}}  & {\makecell{Total\\(GPU h) $\downarrow$}} & PSNR$\uparrow$ & FVD$\downarrow$ \\
\midrule
\multirow{4}{*}{\textbf{CMR}}
& w/o TSP          & 48.01 & 2008.23 & 480.7 & 535.1 & 25.76 & 1982.53 \\
& w/o AAP          & 48.56 & 2011.48 & 483.0 & 536.7 & 27.25 & 1627.98 \\
& w/o TSP \& AAP   & 48.01 & 2008.23 & 480.6 & 534.0 & 25.31 & 2376.82 \\
& \cellcolor{tablepurple!70}\textbf{CardioFAD (Full)} & \cellcolor{tablepurple!70}48.56 & \cellcolor{tablepurple!70}2011.48 & \cellcolor{tablepurple!70}483.2 & \cellcolor{tablepurple!70}536.9 & \cellcolor{tablepurple!70}\textbf{29.00} & \cellcolor{tablepurple!70}\textbf{1291.87} \\
\midrule

\multirow{4}{*}{\textbf{ECHO}}
& w/o TSP          & 44.45 & 247.65 & 202.3 & 224.8 & 27.13 & 368.04 \\
& w/o AAP          & 45.00 & 249.19 & 203.9 & 226.6 & 27.87 & 331.53 \\
& w/o TSP \& AAP   & 44.45 & 247.65 & 201.6 & 224.0 & 26.56 & 428.77 \\
& \cellcolor{tablepurple!70}\textbf{CardioFAD (Full)} &  \cellcolor{tablepurple!70}45.00 & \cellcolor{tablepurple!70}249.19 &  \cellcolor{tablepurple!70}204.0 & \cellcolor{tablepurple!70}226.7 &  \cellcolor{tablepurple!70}\textbf{28.93} & \cellcolor{tablepurple!70}\textbf{239.28} \\
\bottomrule
\end{tabular}
}
\label{Wall-clock_Cost}
\end{table*}

\clearpage
\section*{NeurIPS Paper Checklist}

\begin{enumerate}

\item {\bf Claims}
    \item[] Question: Do the main claims made in the abstract and introduction accurately reflect the paper's contributions and scope?
    \item[] Answer: \answerYes{} 
    \item[] Justification: The main claims presented in our abstract and introduction accurately reflect the contribution of our paper.

\item {\bf Limitations}
    \item[] Question: Does the paper discuss the limitations of the work performed by the authors?
    \item[] Answer: \answerYes{} 
    \item[] Justification: We discuss the limitations of our work in Sec.~\ref{sec:conclusion}. 

\item {\bf Theory assumptions and proofs}
    \item[] Question: For each theoretical result, does the paper provide the full set of assumptions and a complete (and correct) proof?
    \item[] Answer: \answerNA{} 
    \item[] Justification: This paper focuses on proposing a keyframe-to-sequence diffusion framework for one-shot cardiac dynamic synthesis, which does not contain theoretical results.

    \item {\bf Experimental result reproducibility}
    \item[] Question: Does the paper fully disclose all the information needed to reproduce the main experimental results of the paper to the extent that it affects the main claims and/or conclusions of the paper (regardless of whether the code and data are provided or not)?
    \item[] Answer: \answerYes{} 
    \item[] Justification: We provide comprehensive experimental and implementation details, as well as hyperparameter settings in Appendix~\ref{app:imple}.

\item {\bf Open access to data and code}
    \item[] Question: Does the paper provide open access to the data and code, with sufficient instructions to faithfully reproduce the main experimental results, as described in supplemental material?
    \item[] Answer: \answerNo{} 
    \item[] Justification: Some datasets used in this study are not publicly available.

\item {\bf Experimental setting/details}
    \item[] Question: Does the paper specify all the training and test details (e.g., data splits, hyperparameters, how they were chosen, type of optimizer) necessary to understand the results?
    \item[] Answer:  \answerYes{} 
    \item[] Justification: We have included the implementation details and the hyperparameters in Sec.~\ref{sec:exp_setting} and Appendix~\ref{app:imple}.

\item {\bf Experiment statistical significance}
    \item[] Question: Does the paper report error bars suitably and correctly defined or other appropriate information about the statistical significance of the experiments?
    \item[] Answer: \answerYes{} 
    \item[] Justification: We report mean and standard deviation over three independent runs in Appendix~\ref{app:multi-seed}. 

\item {\bf Experiments compute resources}
    \item[] Question: For each experiment, does the paper provide sufficient information on the computer resources (type of compute workers, memory, time of execution) needed to reproduce the experiments?
    \item[] Answer: \answerYes{} 
    \item[] Justification: We have provided the computing cost in Appendix~\ref{app:efficiency}.
    
\item {\bf Code of ethics}
    \item[] Question: Does the research conducted in the paper conform, in every respect, with the NeurIPS Code of Ethics \url{https://neurips.cc/public/EthicsGuidelines}?
    \item[] Answer: \answerYes{} 
    \item[] Justification: We have read and understood the NeurIPS Code of Ethics and have done our best to conform.

\item {\bf Broader impacts}
    \item[] Question: Does the paper discuss both potential positive societal impacts and negative societal impacts of the work performed?
    \item[] Answer: \answerNA{} 
    \item[] Justification: There is no potential societal impact of our work.
    
\item {\bf Safeguards}
    \item[] Question: Does the paper describe safeguards that have been put in place for responsible release of data or models that have a high risk for misuse (e.g., pre-trained language models, image generators, or scraped datasets)?
    \item[] Answer: \answerYes{} 
    \item[] Justification: Our experiments do not involve this aspect.

\item {\bf Licenses for existing assets}
    \item[] Question: Are the creators or original owners of assets (e.g., code, data, models), used in the paper, properly credited and are the license and terms of use explicitly mentioned and properly respected?
    \item[] Answer: \answerYes{} 
    \item[] Justification: We cite the original paper that produced the code package or dataset.

\item {\bf New assets}
    \item[] Question: Are new assets introduced in the paper well documented and is the documentation provided alongside the assets?
    \item[] Answer: \answerNA{} 
    \item[] Justification: The paper does not release new assets.

\item {\bf Crowdsourcing and research with human subjects}
    \item[] Question: For crowdsourcing experiments and research with human subjects, does the paper include the full text of instructions given to participants and screenshots, if applicable, as well as details about compensation (if any)? 
    \item[] Answer: \answerNA{} 
    \item[] Justification: This paper does not involve this aspect.

\item {\bf Institutional review board (IRB) approvals or equivalent for research with human subjects}
    \item[] Question: Does the paper describe potential risks incurred by study participants, whether such risks were disclosed to the subjects, and whether Institutional Review Board (IRB) approvals (or an equivalent approval/review based on the requirements of your country or institution) were obtained?
    \item[] Answer: \answerNA{} 
    \item[] Justification: This paper does not involve this aspect.

\item {\bf Declaration of LLM usage}
    \item[] Question: Does the paper describe the usage of LLMs if it is an important, original, or non-standard component of the core methods in this research? Note that if the LLM is used only for writing, editing, or formatting purposes and does \emph{not} impact the core methodology, scientific rigor, or originality of the research, declaration is not required.
    \item[] Answer: \answerNA{} 
    \item[] Justification: This work does not incorporate any LLMs as part of the core methodology. LLMs were used only for writing and editing suggestions in this paper.

\end{enumerate}

\end{document}